\documentclass[usenatbib]{mn2e}

\usepackage{epsfig}
\usepackage{amssymb}
\usepackage{url}
\usepackage{color}
\newcommand{\new}[1]{\textcolor{black}{#1}}

\newcommand{\lumin}{\mathrm{L}}
\newcommand{\halo}{\mathrm{h}}
\newcommand{\disc}{\mathrm{d}}
\newcommand{\bulge}{\mathrm{b}}
\newcommand{\peri}[1]{\mathrm{p#1}}
\newcommand{\apo}[1]{\mathrm{a#1}}
\newcommand{\tail}[1]{\mathrm{tail#1}}
\newcommand{\vect}[1]{\mathbf{#1}}

\title[Encounter dynamics and tidal response]{Transformations of galaxies.
III. Encounter dynamics and tidal response as functions of galaxy structure}
\author[J.E. Barnes]{Joshua E. Barnes$^{1,2}$\\
$^1$Institute for Astronomy, University of Hawaii, 2680 Woodlawn Drive,
Honolulu, HI 96822, USA\\
$^2$Yukawa Institute for Theoretical Physics, Kyoto University,
Sakyo-ku, Kyoto 606-8502 Japan}

\begin{document}

\maketitle

\begin{abstract}
Tidal interactions between disc galaxies depend on galaxy structure,
but the details of this relationship are incompletely understood. I
have constructed a three-parameter grid of bulge/disc/halo models
broadly consistent with $\Lambda$CDM, and simulated an extensive
series of encounters using these models. Halo mass and extent strongly
influence the dynamics of orbit evolution. In close encounters, the
transfer of angular momentum mediated by the dynamical response of
massive, extended haloes can reverse the direction of orbital motion
of the central galaxies after their first passage. Tidal response is
strongly correlated with the ratio $v_\mathrm{e} / v_\mathrm{c}$ of
escape to circular velocity within the participating discs.  Moreover,
the same ratio also correlates with the rate at which tidal tails are
reaccreted by their galaxies of origin; consequently, merger remnants
with `twin tails', such as NGC~7252, may prove hard to reproduce
unless $(v_\mathrm{e} / v_\mathrm{c})^2 \lesssim 5.5$. The tidal
morphology of an interacting system can provide useful constraints on
progenitor structure. In particular, encounters in which halo dynamics
reverses orbital motion exhibit a distinctive morphology which may be
recognized observationally.  Detailed models attempting to reproduce
observations of interacting galaxies should explore the likely range
of progenitor structures along with \new{other} encounter parameters.
\end{abstract}


\section{INTRODUCTION}

Numerical studies of interacting galaxies have come far since the
first efforts, but every study to date has grappled with the huge
variety of encounters which are possible.  In a fundamental paper,
\citet[hereafter TT]{TT1972} introduced a key set of parameters and
illustrated how these parameters influence the outcome of a tidal
encounter.  Many subsequent studies have adopted similar
parameterizations and explored some subset of galactic encounters.
For example, a number of authors have focused on equal-mass
encounters, in which both galaxies have the same initial structure,
and systematically examined the tidal response as a function of
encounter parameters such as initial disc orientation.

Somewhat less attention has been given to the ways in which tidal
interactions depend on the \textit{internal} structures of the
galaxies involved.  For TT, this issue was almost moot -- with all the
gravitating mass concentrated in a central point, all they needed to
do was to vary the outer radii of their test-particle discs.  Once
self-consistent simulations became possible \citep[e.g.,][]{W1978,
FS1982, B1988}, the question of internal structure, and in particular
the amount of dark matter, became more interesting.  \citet{DMH1996}
simulated encounters between galaxies with various disc/halo mass
ratios, and reported that only relatively low-mass haloes permitted
the formation of long tidal tails.  Later studies \citep[hereafter
SW]{MDH1998, DMH1999, B1999, SW1999} examined this claim in detail,
and converged on a more precise conclusion: tail length constrains
potential-well \textit{shape}, not disc/halo fraction (see
\S~\ref{sec:tailform}).  But these studies largely focused on the
specific question of tail formation, and paid less attention to the
broader relationship between internal structure and tidal
interactions.

Hierarchical galaxy formation includes stochastic elements which
inevitably produce variations in galactic properties and structure.
Disc galaxies display a range of rotation curves; some have gently
rising profiles, while others reach a maximum at a few disc scale
lengths and remain flat or gradually decline at larger radii
\citep[e.g.,][]{CvG1991, dBBTOK2008}.  Even within the constraints
implied by the baryonic Tully-Fischer relationship
\citep[e.g.][]{F1999, MSBdB2000, V2001} there seems to be room for
interesting variety, with the luminous mass fraction (including both
stars and gas) varying by a factor of a few from galaxy to galaxy
\citep{Z+2014}.  Thus in modeling tidal encounters of disc galaxies,
there are good reasons to experiment with different halo structures.

Ultimately, detailed modeling of interacting galaxies may provide a
way to study the initial structure of their progenitors.  This could
offer a unique window on halo properties, involving the dark matter as
a fully dynamical participant and probing its distribution out to
relatively large radii.  At present, however, it's not clear how much
information can be extracted from modeling; for example, can dynamical
modeling correctly diagnose halo structure, or do the plethora of
parameters available insure that the observations can be matched
tolerably well even with the wrong choice of halo model?  To frame
this question more tightly, suppose we \textit{knew} the orbital and
viewing geometry of an interacting system -- under this very idealized
set of circumstances, \new{can} detailed modeling yield a correct
picture of halo structure?

\subsection{Tail formation}
\label{sec:tailform}

Distinctive tidal features form when galactic discs are subject to
relatively brief but violent tides during close encounters (TT).
Unlike the rather broad tidal bulges that the Moon creates on Earth,
the tidal features extracted from spinning galactic discs are often
quite narrow and extended.  In a typical encounter, each disc produces
two such features: a \textit{bridge} toward the galaxy's companion,
and a \textit{tail} stretching in the opposite direction.  The most
pronounced bridges and tails develop when a galaxy's spin angular
momentum is parallel to the orbital angular momentum of a passing
companion.  In this case, the angular velocities of individual stars
resonate with the angular velocity of the companion, creating a strong
response (TT; \citealt{DOVFGH2010}).  An extended tail can only
form if the companion's mass is comparable to or exceeds the victim's;
otherwise, the result is a weaker `counter-arm' which soon falls
back into the parent galaxy.  In systems like NGC~4676 where
\textit{both} galaxies exhibit tails, this implies that the two
galaxies have comparable masses.

SW, \new{building on earlier work by \citet{MMW1998} and
\citet{B1999}}, \marginpar{\new{[1]}}
showed that even under otherwise favorable
circumstances, \new{discs don't produce long tidal tails} if the
quantity
\begin{equation}
  \mathcal{E} =
    \frac{v_\mathrm{e}^2}{v_\mathrm{c}^2}
  \label{eq:sw-escape-param}
\end{equation}
exceeds a certain value.  \new{Here} $v_\mathrm{e}$ and $v_\mathrm{c}$
are the escape and circular velocity, respectively, \new{evaluated
near the disc's half mass radius; thus $\mathcal{E}$ is the ratio of
the escape energy to the kinetic energy for a circular orbit}.
\new{Only discs with $\mathcal{E} \lesssim 6.5$ readily form tails in
equal-mass encounters.}  The situation for encounters with mass ratios
$\mu \ne 1$ has not yet been investigated.

\begin{figure}
\begin{center}
\epsfig{file=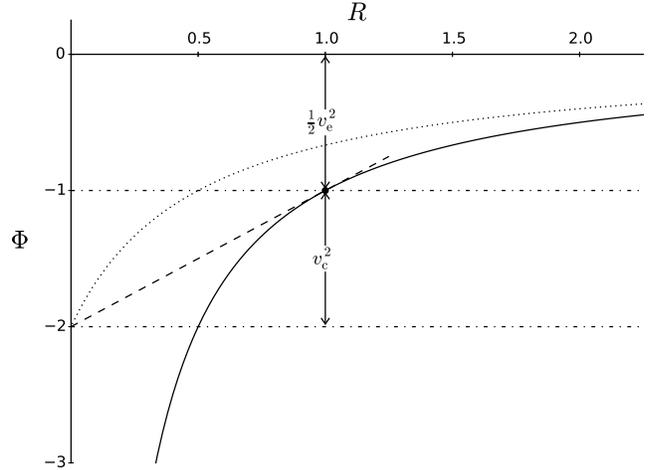,width=\columnwidth,
bbllx=43,bblly=219,bburx=538,bbury=583}
\caption{Relationship between escape velocity $v_\mathrm{e}$ and
circular velocity $v_\mathrm{c}$. The solid curve shows a $\Phi =
-1/R$ potential.  The dashed line, tangent to the curve $\Phi(R)$ at
radius $R$, intercepts the $R = 0$ axis at $\Phi = \Phi(R) -
v_\mathrm{c}^2$. \new{Dot-dash lines at $\Phi = -1$ and $-2$ help to
show that $\frac{1}{2} v_\mathrm{c}^2 = v_\mathrm{e}^2$ for a
Keplerian potential.}  The dotted curve \new{is} a \citet{H1990}
potential, $\Phi = -1 / (R + a)$, with scale radius $a = \new{0.5}$;
this is simply the $-1/R$ curve translated to the left by $a$.
\label{fig:escplot}}
\end{center}
\end{figure}

This criterion can be directly visualized on a plot of gravitational
potential $\Phi(R)$.  Fig.~\ref{fig:escplot} illustrates the
relationship between $v_\mathrm{e}^2 = -2 \Phi(R)$ and $v_\mathrm{c}^2
= R \, d \Phi / d R$.  For a Keplerian potential, $v_\mathrm{e}^2 = 2
v_\mathrm{c}^2$ everywhere, so $\mathcal{E}_\mathrm{Kep} = 2$.  This
implies that \new{test-particle discs in Keplerian potentials can
easily form tidal tails, as TT originally demonstrated}.  For
comparison, consider the \citet{H1990} model, which has a potential of
the form $\Phi(R) \propto 1 / (R + a)$, where $a > 0$ is the scale
radius.  It's evident that $v_\mathrm{e}^2 > 2 v_\mathrm{c}^2$
everywhere, and $\mathcal{E}\new{(R)} = 2 (R + a) / R$,
\new{interpreted just for this paragraph as a function of $R$},
\new{becomes arbitrarily large} in the $R \ll a$ limit.  The condition
$\mathcal{E} \lesssim 6.5$ is satisfied at radii $R \gtrsim 0.44 a$.
Thus, to produce tails, a disc of test particles embedded in a
\citeauthor{H1990} potential with scale radius $a$ must have a median
radius $R_{1/2} \gtrsim 0.44 a$.  Similar considerations apply to
other extended mass distributions, \new{including those with discs of
finite mass}. \new{In general, $\mathcal{E}(R)$ approaches the
Keplerian value $\mathcal{E}_\mathrm{Kep} = 2$ from above as $R \to
\infty$.  This implies that more extended discs have smaller values of
$\mathcal{E}$, and therefore form tails more easily.}

This simple example suggests that modeling will indeed be able to
provide some information on halo structure -- the very existence of
long tidal tails implies, at a minimum, that some galaxies have
\new{discs with} $\mathcal{E} \lesssim 6.5$.  But further work is
needed to find out how much more we can learn.

\subsection{Outline}

To determine what detailed modeling of tidal encounters might teach
us, it seems reasonable to investigate encounters between a variety of
galaxy models.  This is not an entirely new direction; in particular,
SW and \citet{DMH1999} provide significant precedents.  The present
study expands on their work, testing a wide range of galaxy models,
classifying tidal responses, and systematically comparing tidal
\new{features}.  Like these studies, it shares one key limitation: the
two galaxies in each encounter are `twins', with the same mass
\textit{and} the same internal structure.  However, a fairly wide
range of internal structures are employed, producing a variety of
interaction dynamics and tidal configurations.

This paper is organized as follows.  Section~2 first develops a set of
galaxy models and identifies those which are stable and therefore
suitable raw material for further investigations, and next describes
the set of encounters simulated using these galaxies.  Section~3
presents simulation results, focusing on orbital evolution, while
Section~4 covers characterization of tidal features.  Section~5 takes
up the hypothetical modeling problem just described, and asks how well
tidal response can constrain halo structure.  Conclusions appear in
Section~6.  Technical details are described in Appendix~A, while tests
of isolated galaxy models appear in Appendix~B.

\section{INITIAL CONDITIONS}

\subsection{Galaxy Models}
\label{sec:galmod}

Each galaxy model contains three collisionless\footnote{Interstellar
material is not included in these simulations.  Gas and stars
generally follow similar trajectories in extended tidal features, and
the added computational expense would be prohibitive.} components,
\new{initialized with explicit density} profiles; in order of
decreasing mass, \new{these are the} halo, disc, and bulge.  The halo
is composed of dark matter, while the disc and bulge are composed of
luminous material (stars), but all are assumed to obey the same
$N$-body equations of motion.

\textbf{1.} The halo has a \citet[][hereafter NFW]{NFW1996, NFW1997}
profile, parameterized by a total mass $M_\halo$, a scale radius
$a_\halo$, and a taper radius $b_\halo$.  Within $R \le b_\halo$, the
profile is
\begin{equation}
  \rho_\halo(R) =
    \frac{M_\halo(a_\halo)}
         {4 \pi (\ln(2) - \frac{1}{2}) R (a_\halo + R)^2}
    \, ,
  \label{eq:halo-profile}
\end{equation}
where $M_\halo(a_\halo)$ is the mass within $a_\halo$. For $R >
b_\halo$, the density profile tapers off exponentially, using the
functional form devised by SW.  The taper radius $b_\halo$ is usually
called the `virial radius' and identified with $R_{200}$, the radius
within which the average density of the halo is $200$ times the
critical density of the universe (NFW).  This identity constrains the
scaling of numerical models to physical units, but isn't directly
relevant for the present calculations, which treat the tapered NFW
profile as a convenient functional form.  In addition, the NFW profile
is used `as is', without adiabatic compression; this is further
discussed in \S~\ref{sec:halo_compress}.

\textbf{2.} The disc has an exponential-isothermal profile,
parameterized by mass $M_\disc$, inverse scale length
$\alpha_\disc$, and scale height $z_\disc$:
\begin{equation}
  \rho_\disc(R,z) =
    \frac{M_\disc}{4 \pi \alpha_\disc^2 z_\disc} \,
    e^{-R \alpha_\disc} \, \mathrm{sech}^2(z / z_\disc) \, .
  \label{eq:disc-profile}
\end{equation}
No outer limit is imposed on the disc profile.  In the present
experiments, discs account for $75$~percent of the luminous material.
The scale height is independent of $R$ and is fixed at $z_\disc =
0.125 / \alpha_\disc$ for all models.

\textbf{3.} The central bulge has a \citet{J1983} profile, parameterized by
mass $M_\bulge$ and scale radius $a_\bulge$:
\begin{equation}
  \rho_\bulge(R) =
    \frac{M_\bulge a_\bulge}{4 \pi R^2 (a_\bulge + R)^2} \, .
  \label{eq:bulge-profile}
\end{equation}
Bulges account for the other $25$~percent by mass of the luminous
material, so $M_\bulge = \frac{1}{3} M_\disc$.  The bulge scale radius
is taken to be $a_\bulge = 0.16 a_\halo$ in all experiments.  Such a
compact bulge has little direct effect on the dynamics of tidal
interactions, but it helps to stabilize the disc against bar
instabilities.  In $N$-body simulations the $r^{-4}$ tail of the bulge
profile presents some difficulties, since the outermost body has
radius $\sim N a_\bulge$; as described in \citet{B2012}, it's
convenient to smoothly taper (\ref{eq:bulge-profile}) at large $R$.

\begin{figure*}
\begin{center}
\epsfig{file=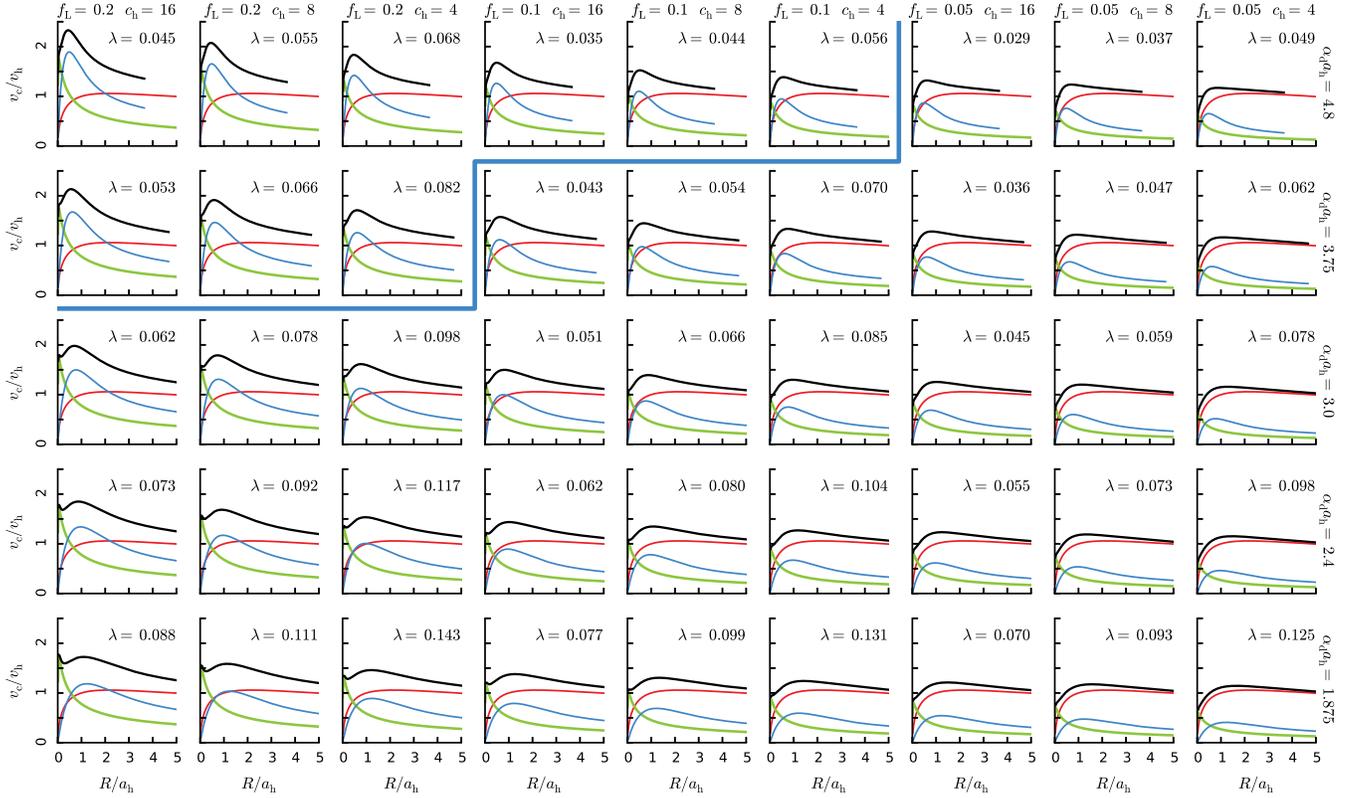,height=\textwidth,angle=270,
bbllx=139,bblly=105,bburx=487,bbury=690}
\caption{Rotation curves for the B/D/H galaxy models.  \new{Each panel
shows the rotation curve for a different galaxy model, with the
luminous fraction $f_\lumin$ and halo concentration $c_\halo$
specified along the top, and the disc compactness, $\alpha_\disc
a_\halo$, specified along the right.}  Black curves are total circular
velocities, while the green, blue, and red curves show the
contributions of the bulges, discs, and haloes, respectively.  Radii
are normalized to $a_\halo$, the halo's scale radius, while velocities
are normalized to $v_\halo \new{= \sqrt{G M_\halo(a_\halo) /
a_\halo}}$, the halo's circular velocity at $R = a_\halo$.  Estimated
values of $\lambda$ are calculated from (\ref{eq:lambda-protogalaxy}).
Models above the heavy blue line are \new{generally too} unstable
\new{to be used for encounter simulations}.
\label{fig:vrot}}
\end{center}
\end{figure*}

\new{The galaxy models used in this paper form a three-dimensional
grid.  Fig.~\ref{fig:vrot} presents circular velocity profiles for the
full set of $3 \times 3 \times 5 = 45$ bulge/disc/halo models
considered here.  In this and subsequent figures, this grid is laid
out in two dimensions, with the ratio of halo mass within $a_\halo$ to
total luminous mass increasing from left to right, and the radial
scale of the disc relative to the halo increasing from top to bottom.
This grid contains a wide variety of models, including some which may
fall outside the gamut of real galaxies.  The rest of this section
tries to place these models in the context of recent descriptions of
galaxy formation in $\Lambda$CDM cosmologies.}

The first and \new{arguably the} most fundamental parameter is the
luminous mass fraction, $f_\lumin = (M_\bulge + M_\disc) / (M_\bulge +
M_\disc + M_\halo)$.  In $\Lambda$CDM models \new{consistent with}
\textit{WMAP} and \textit{Planck}, baryons comprise $16 \pm 1$~percent
of the matter \citep{H+2013, P+2015}.  If all \new{the} baryons
\new{in a proto-galaxy} were incorporated into \new{a} galactic disc
and bulge then it would be appropriate to set $f_\lumin \simeq 0.16$.
This value is almost certainly too high in view of the evident
inefficiency of galaxy formation, as illustrated by observations of
massive outflows from star-forming galaxies \citep[e.g.][]{P+2001,
SE+2010, M+2012} and the large reservoirs of gas in rich galaxy
clusters \citep[e.g.][]{G+2009}.  To explore trends with luminous
fraction, experiments are run with $f_\lumin = 0.2$, $0.1$, and
$0.05$.  \new{The high end of this range, which is beyond the
canonical value of $0.16$, is included to make contact with earlier
experiments \citep[e.g.,][]{B1988, B1992}.  At the low end, $f_\lumin
= 0.05$ allows the luminous discs to retain some degree of
self-gravity; still lower values, although astrophysically possible,
\new{effectively} relegate the discs to test-particle status.}

The second parameter is the halo concentration, $c_\halo = b_\halo /
a_\halo$.  In $\Lambda$CDM simulations, the concentration of a halo
depends on its formation history; haloes which have recently been
restructured by major mergers typically have low concentrations, while
those which have been quietly accreting for a long time have higher
concentrations \citep[e.g.,][]{ZJMB2003, L+2014}.  As noted above,
(\ref{eq:halo-profile}) is used here as a convenient function, and the
value of $b_\halo$ is not tied directly to the cosmology.  To sample a
range of both realistic and counterfactual possibilities, values of
$c_\halo = 4$, $8$, and $16$ are adopted; the first of these makes
contact with earlier experiments which typically used rather small
haloes.

The third parameter is the disc `compactness', $\alpha_\disc a_\halo$
(the slightly nonstandard terminology is \new{intended as} a reminder
that larger values of $\alpha_\disc a_\halo$ imply \textit{smaller}
discs, and vice versa).  Values of $\alpha_\disc a_\halo = 1.875$,
$2.4$, $3.0$, $3.75$, and $4.8$ are used here.  Note that these values
are almost equally spaced logarithmically by factors very close to
$\sqrt[3]{2}$.  The range of $\alpha_\disc a_\halo$ values
\new{adopted here} is dictated by two considerations.

\new{First, not all of the models in Fig.~\ref{fig:vrot} are stable.
In particular, the disc-dominated models at the upper left of the grid
rapidly develop strong bars.  Appendix~B describes stability tests for
these galaxy models which set an upper limit on $\alpha_\disc a_\halo$
for a given choice of $f_\lumin$ and $c_\halo$.}

Second, very extended discs require large amounts of angular momentum.
\new{The angular momentum of a proto-galaxy is} quantified by the
dimensionless spin parameter
\begin{equation}
  \lambda =
    \frac{J_\mathrm{pg} \sqrt{-E_\mathrm{pg}}}{G M_\mathrm{pg}^{5/2}} \, .
  \label{eq:lambda-protogalaxy}
\end{equation}
Here $M_\mathrm{pg}$, $J_\mathrm{pg}$, and $E_\mathrm{pg}$ are the
proto-galaxy's mass, angular momentum, and binding energy,
respectively.  In a simple picture \new{of galaxy formation} where
discs form via gradual gas cooling within initially well-mixed and
undifferentiated haloes \citep[e.g.,][]{FE1980, F1983, DSS1997,
MMW1998}, these parameters may be estimated as follows.  Neglect of
accretion or outflows implies $M_\mathrm{pg} = M_\bulge + M_\disc +
M_\halo$.  Gas and dark matter are both subject to the same tidal
torques \citep{H1949, P1969}, and therefore should have the same
specific angular momenta; assuming that the gas conserves angular
momentum as it cools, $J_\mathrm{pg} / M_\mathrm{pg} = J_\disc /
M_\disc$, where the right-hand side refers to the \textit{present}
disc.  Finally, $E_\mathrm{pg}$ may be \new{computed} by assuming that
the virialized proto-galaxy had the same radial
distribution\footnote{\citet{MMW1998} \new{and SW} take adiabatic
compression into account \new{in computing $E_\mathrm{pg}$}, but this
is a relatively small correction.} as the present halo.

Fig.~\ref{fig:vrot} shows $\lambda$ values for each model,
\new{estimated using (\ref{eq:lambda-protogalaxy}).  In each column of
this figure, $\lambda$ scales in rough proportion to
$\alpha_\disc^{-1}$; larger discs have more angular momentum.}
Simulations of \new{structure} formation in $\Lambda$CDM indicate that
the spin parameter has a median value $\lambda_\mathrm{med} \simeq
0.034$, with some dependence on the algorithm used to define bound
haloes \citep[e.g.,][]{B+2007}; the distribution of $\lambda$ is
\new{rather wide}, with 10th and 90th percentiles differing by a
factor of $\sim 5$ \citep{MvdBW2010}.  Most of the stable galaxy
models in Fig.~\ref{fig:vrot} have estimated $\lambda$ values
exceeding $\lambda_\mathrm{med}$.  \new{However}, given the width of
the $\lambda$ distribution, it seems reasonable to view models with
$\lambda \lesssim 0.08$ as generally consistent with \new{simple
pictures of} galaxy formation in a $\Lambda$CDM universe
\new{\citep[e.g.,][]{MMW1998}}.

\new{Nearly two-thirds of the models in the bottom two rows of
Fig.~\ref{fig:vrot} have estimated $\lambda$ values exceeding $0.08$.
These models will be retained as a hedge against the possibility that
the simple picture of galaxy formation invoked above is incomplete.
For example, outflows may preferentially eject gas with low angular
momentum \citep{B+2011, G+2015}, leaving material with high angular
momentum to form discs; alternately, accretion via cold flows may
introduce gas with high angular momentum \citep{SBBMDWM2013} which can
build up larger discs.  Retaining these models yields a total of $36$
stable galaxy models\footnote{\new{None of these models reproduce the
monotonically rising rotation curves observed in many low-mass and
low-surface-brightness disc galaxies; luminous fractions $f_\lumin <
0.025$ appear necessary to obtain such curves using NFW halos,
although larger $f_\lumin$ values are possible if halos with shallower
central profiles are used.}} which will be used for encounter
simulations.}

\subsubsection{Halo compression}
\label{sec:halo_compress}

Unlike earlier studies (e.g., \citealt{MMW1998}; SW), the NFW halo
profiles in the present models were not modified to account for
adiabatic compression by the gravitational field of the disc and
bulge.  This is partly a matter of convenience; the process of model
construction and any auxiliary calculations are more straightforward
if the NFW profile is used without modification.  However, there are
two additional considerations.

\textbf{1.} The standard halo compression algorithm, due to
\citet{BFFP1986}, is based on the assumption that the halo responds as
if its constituent particles are on circular orbits.  In practice,
this may not be a good assumption; haloes formed by gravitational
collapse are likely to have \textit{radially} biased velocity
distributions.  A number of studies \citep{B1987, S1999, WK2002,
GKKN2004, SM2005, TWPS2010} have found that the \citet{BFFP1986}
algorithm significantly \textit{overestimates} the response of
initially isotropic or radially-biased haloes.  \citet{SM2005}
describe an algorithm, based on \citet{Y1980}'s treatment of adiabatic
compression in spherical systems, which describes the compression of
such haloes more accurately.

\textbf{2.} Observational evidence suggests that the effect of galaxy
formation on halo structure is considerably more complex than
models of adiabatic compression imply.  Dwarf disc galaxies, in
particular, appear to have haloes with constant-density cores or cusps
shallower than the $r^{-1}$ NFW profile \citep[e.g.,][]{CCF2000,
dBMR2001, dBBM2003, SMvdBB2003}.  One possible explanation invokes
what might be called \textit{non-adiabatic decompression} of dark
haloes in response to explosive ejection of baryonic material
\citep[e.g.,][and references therein]{G+2012}.  While compressed NFW
haloes fit the rotation curves of massive galaxies fairly well
\citep[e.g.,][]{SM2005, D+2011}, direct evidence for massive outflows
\citep[e.g.][]{P+2001, SE+2010, M+2012} shows that baryons don't always
accumulate in a gradual and monotonic fashion.

In sum, the standard recipe for halo compression should probably be
replaced by a more accurate treatment including both adiabatic and
non-adiabatic processes.  However, it's not yet clear what effects
must be included.  The exploratory calculations presented here are
relatively insensitive to the details of the inner halo profile;
\new{including} halo compression \new{would not} alter the main
results of this study.

\subsubsection{Initialization}
\label{sec:init}

The bulge, disc, and halo components of each model are initialized in
approximate dynamical equilibrium with their combined gravitational
field.  For the halo and bulge, a smoothing formalism \citep{B2012} is
used to compute their contributions to the gravitational field, while
the disc's contribution is approximated by an equivalent spherical
mass distribution.  Isotropic distribution functions for the bulge and
halo are computed using Eddington's formula, and \new{sampled to
obtain position and velocity coordinates} \citep[e.g.,][]{B2012}.  The
disc is initialized using Jeans' equations to constrain moments of the
velocity distribution \citep[e.g.,][]{BH2009}.  While this procedure
is somewhat ad hoc, it's very fast; this is \new{an advantage} when
many simulations are planned.  The large number of experiments
dictates relatively modest particle numbers: $N_\bulge = 16384$,
$N_\disc = 49152$, and $N_\halo = 65536$ to~$311296$.  However,
\new{the simulations are large enough} to study tidal responses.
Further details of the simulations are given in Appendix~A.

\subsection{Encounter Survey}

All of the encounters described here have the same mass ratio, $\mu =
1$, initial orbital eccentricity, $e = 1$, and encounter geometry.
\new{The two galaxy models in each encounter have identical
parameters.}  \new{One} disc lies exactly in the orbital plane and
rotates in the same direction that the two galaxies pass each other;
this disc therefore has inclination $i_1 = 0^\circ$.  The other disc
has an inclination of $i_2 = 71^\circ$ and a nominal pericentric
argument, relative to the idealized Keplerian orbit, of $\omega_2 =
+30^\circ$.  Thus, while both discs have prograde ($i < 90^\circ$)
encounters, the second disc is tilted by a fairly large angle,
generating rather different tidal features.

The primary \new{encounter survey} spans a grid of four parameters.
Three describe the galaxy model and \new{were introduced in
\S~\ref{sec:galmod}}. The remaining parameter specifies the
pericentric separation of the initial orbit, $r_\peri{} / a_\halo$.
All four of these parameters are dimensionless quantities; to
summarize, the values used are
\begin{equation}
  \begin{array}%
        {r@{\quad}c@{\quad}l@{\quad}l@{\quad}l@{\quad}l@{\quad}l@{\quad}}
    f_\lumin & = & 0.2, & & 0.1, & & 0.05, \\
    c_\halo & = & 16,   & & 8,   & & 4, \\
    \alpha_\disc a_\halo & = & 1.875, & 2.4, & 3.0, & 3.75, & 4.8, \\
    r_\peri{} / a_\halo & = & 0.5, & & 1.0, & & 2.0 \, . \\
  \end{array}
  \label{eq:primary-encounter-grid}
\end{equation}
Since only $36$ of the $45$ galaxy models are stable, the primary
sample contains a total of $108$ different encounters.

In addition to the primary sample, $24$ encounters with pericentric
separations interpolating between the values in
(\ref{eq:primary-encounter-grid}) were run.  This secondary sample
contains six galaxy models; three with luminous fraction $f_\lumin
= 0.1$, disc compactness $\alpha_\disc a_\halo = 3.0$, and
halo concentration $c_\halo = 16, 8, 4$, and three with
$f_\lumin = 0.05$, $\alpha_\disc a_\halo = 4.8$, and
$c_\halo = 16, 8, 4$.  Pericentric separations of
\begin{equation}
  r_\peri{} / a_\halo = 0.625, 0.8, 1.25, 1.6 \, ,
  \label{eq:secondary-peri-grid}
\end{equation}
when combined with the primary grid, provide finer coverage in
$r_\peri{} / a_\halo$.  This secondary sample is useful in exploring
trends with pericentric separation for the six models it includes.

An encounter's pericentric separation $r_\peri{}$ is related to the
angular momentum $J_\mathrm{orb}$ of its initial parabolic orbit:
\begin{equation}
  J_\mathrm{orb} = \sqrt{G M_\mathrm{g}^3 r_\peri{}} \, ,
  \label{eq:angular-momentum-orbit}
\end{equation}
where $M_\mathrm{g}$ is the mass of a single galaxy.  This angular
momentum is presumably generated by tidal torques acting on the two
galactic haloes as they collapse out of the Hubble flow, reach their
maximum separation, and fall back towards each other; the amount of
momentum torques generate implies an upper limit to $r_\peri{}$.  To
estimate this limit, assume that haloes merge without significant
ejection of mass, angular momentum, or binding energy; the remnant
will then have spin parameter
\begin{equation}
  \lambda_\mathrm{orb} =
    \frac{J_\mathrm{orb}\sqrt{-E_\mathrm{orb}}}
         {G (2 M_\mathrm{g})^{5/2}} \, ,
  \label{eq:lambda-orbit}
\end{equation}
where $E_\mathrm{orb}$ is binding energy of the initial configuration.
Since initially the orbit is parabolic and the galaxies are
well-separated, $E_\mathrm{orb}$ is just twice the binding energy of a
single galaxy, $E_\mathrm{g} = - K G M_\mathrm{g}^2 / a_\halo$,
where the form factor $K$ has a weak dependence on $c_\halo$,
$f_\lumin$, and $\alpha_\disc a_\halo$.  Evaluating this
factor numerically yields
\begin{equation}
  \lambda_\mathrm{orb} \simeq
    (0.067 \pm 0.017) \sqrt{r_\peri{} / a_\halo} \, ,
  \label{eq:lambda-orbit-value}
\end{equation}
where the given uncertainty encompasses the full range of values
possible for all $36$ stable galaxy models.  The upshot is that
encounters with $r_\peri{} / a_\halo = 0.5$ to~$2$ have orbital
angular momenta within the range which can be produced by the tidal
torque mechanism.  Wider encounters are probably rare, requiring
special circumstances to \new{generate} so much angular momentum.  On
the other hand, closer encounters can occur and are certainly worth
investigating.


\section{ORBITAL DYNAMICS}

A tidal encounter between two extended, self-gravitating objects
transfers energy and momentum from relative motion to internal degrees
of freedom.  As a result, the orbits of interacting galaxies evolve
and eventually decay, \new{culminating in a} merger.

\subsection{Encounter characterization}

While Keplerian trajectories neatly parameterize the ingoing orbits of
a pair of initially well-separated galaxies, they don't describe the
circumstances of deeply interpenetrating encounters very well.  In
such encounters, galaxies begin diverging from their initial orbits
even before their first passage.  Such divergence is to be expected:
once they are close enough to interpenetrate, the mutual gravitational
acceleration of two spatially extended structures is less than that of
two equivalent point masses.  With less acceleration to bend their
trajectories, the galaxies undergo a first passage both wider and
slower than their initial Keplerian orbit would imply.

Accurate orbital trajectories are needed to examine this effect.  At
every time-step, the central position $\overline{\vect{r}}_j$ and
velocity $\overline{\vect{v}}_j$ of galaxy $j$ were computed by
averaging over a fixed set $\mathcal{C}_j$ of tightly-bound bodies.
These sets were constructed by initially sorting the bulge bodies of
galaxy $j$ by binding energy, and using the $25$ percent most tightly
bound as $\mathcal{C}_j$.  While some diffusion in binding energy
occurs due to $N$-body scattering and dynamical evolution, the most
tightly-bound quartile of the bulge is stable and provides a robust
determination of galaxy position, capable of tracking the motion of
the dynamical centre through at least the first three pericentric
passages.  (Galaxy velocities are determined a bit less accurately
since the bodies in $\mathcal{C}_j$ have a larger spread in velocity
than in position, but in practice the inner quartile of each bulge
averages over enough bodies to provide good results.)  As these
trajectories are computed, it's straightforward to identify the
instant $t_\peri{1}$ of closest approach; a snapshot of the system at
this time is saved for subsequent analysis.  Let $r_\peri{1} =
|\overline{\vect{r}}_2 - \overline{\vect{r}}_1|$ and $v_\peri{1} =
|\overline{\vect{v}}_2 - \overline{\vect{v}}_1|$ be the separation and
relative velocity at time $t_\peri{1}$; these may be compared to the
corresponding Keplerian values, $r_\peri{}$ and $v_\peri{}$,
respectively.

\begin{figure}
\begin{center}
\epsfig{file=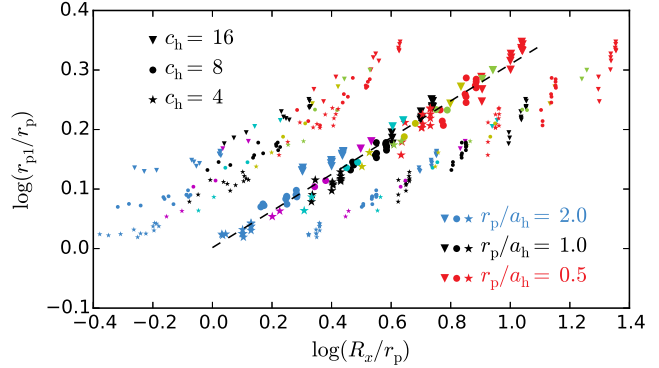,width=\columnwidth,
bbllx=80,bblly=263,bburx=489,bbury=498}
\caption{Scatter plot showing the relationship between the normalized
pericentric separation $r_\peri{1} / r_\peri{}$ and the
interpenetration parameter $R_{x} / r_\peri{}$.  All $132$ encounters
are plotted.  The large symbols show results for $R_{x} = R_{1/2}$,
the half-mass radius, while smaller symbols to the left and right show
results for $R_{x} = R_{1/4}$ and $R_{3/4}$, respectively.  Symbol
type indicates halo concentration, while color indicates initial
pericentric separation (see Fig.~\ref{fig:trakxy} for a complete key).
The dashed line shows a power law fit, with a slope of $0.308$.
\label{fig:interpen}}
\end{center}
\end{figure}

The ratio $R_\mathrm{1/2} / r_\peri{}$, where $R_\mathrm{1/2}$ is the
galactic half-mass radius, quantifies the degree of interpenetration
which \textit{would} occur if the galaxies \new{remained} on their
initial trajectories.  Despite its somewhat hypothetical formulation,
this ratio is a good predictor of orbital behavior
\citep[e.g.,][]{FS1982, B1992}.  For example, it predicts the actual
pericentric separation $r_\peri{1}$; as Fig.~\ref{fig:interpen} shows,
all $132$ simulations presented here follow a fairly tight power-law
of the form $r_\peri{1} / r_\peri{} \simeq (R_\mathrm{1/2} /
r_\peri{})^{0.308}$.  Similar relationships are obtained using
$R_{1/4}$ and $R_{3/4}$, the radii enclosing one quarter and three
quarters of the mass, respectively, although the former correlation
shows more scatter.  In the limit of very wide, non-interpenetrating
passages, $r_\peri{1} / r_\peri{}$ presumably approaches $1$ from
above, so these empirical power-laws can't be universal.  However,
Fig.~\ref{fig:interpen} nicely illustrates how initial pericentric
separation and halo concentration jointly influence first passage via
the degree of interpenetration.  Deviations from Keplerian
trajectories are largest for close encounters (red: $r_\peri{} /
a_\halo = 0.5$) between extended galaxies (triangles: $c_\halo = 16$),
and smallest for wide encounters (blue: $r_\peri{} / a_\halo = 2$)
between compact galaxies (stars: $c_\halo = 4$).

\begin{figure}
\begin{center}
\epsfig{file=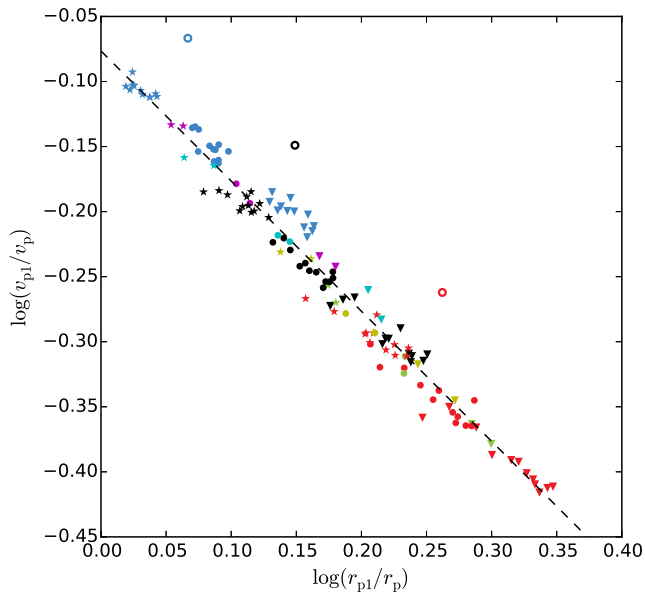,width=\columnwidth,
bbllx=72,bblly=186,bburx=493,bbury=575}
\caption{Scatter plot showing measured pericentric separation
$r_\peri{1}$ and relative velocity $v_\peri{1}$, normalized by
the corresponding values for the initial Keplerian orbit,
$r_\peri{}$ and $v_\peri{}$.  As in Fig.~\ref{fig:interpen},
all $132$ encounters are plotted, symbol type indicates halo
concentration, and color indicates pericentric separation.  In
addition, three open circles represent encounters of heavily softened
`point' masses.  The dashed line shows the relationship
$j_\peri{1} = 0.838 \, j_\mathrm{orb}$.
\label{fig:radvelperi}}
\end{center}
\end{figure}

Fig.~\ref{fig:radvelperi} displays measured pericentric separations
$r_\peri{1}$ and relative velocities $v_\peri{1}$, normalized by the
corresponding values for the initial Keplerian orbits.  This plot
reveals a simple pattern: across the entire set of $132$
self-consistent encounters, the specific angular momentum at
pericentre $j_\peri{1} = r_\peri{1} v_\peri{1} \simeq (0.838 \pm
0.025) \, j_\mathrm{orb}$, where $j_\mathrm{orb} = J_\mathrm{orb} / (2
M_\mathrm{g})$ is the specific angular momentum of the initial orbit.
This factor of $\sim 0.84$ appears because tidal interactions,
operating even \textit{before} first passage, have transferred about
$16$~percent of the initial angular momentum from orbital motion to
internal degrees of freedom within each galaxy.  It seems remarkable,
given the range of encounters studied here, that such a consistent
fraction of orbital angular momentum is lost.  For comparison, the
three open symbols in this figure show experiments with very heavily
softened point masses, which can be thought of as rigid mass profiles.
Because they cannot deform, their orbits do not decay, and their
encounters conserve orbital momentum exactly.

A final point concerns the argument of pericentre $\omega_2$ for the
second ($i_2 = 71^\circ$) discs.  The nominal value of $\omega_2 =
+30^\circ$ would imply that the first ($i_1 = 0^\circ$) galaxy passes
through the plane of this disc some time \textit{before} pericentre.
However, because these self-consistent orbits deviate quite strongly
from their Keplerian counterparts, the actual positions of the two
galaxies at $t_\peri{1}$ typically places the first galaxy's centre
close to the second galaxy's spin plane.  For this sample of
encounters, the second disc's \textit{effective} argument of
pericentre is $\omega_2^\mathrm{eff} \new{{} \simeq 0^\circ \pm
15^\circ}$, instead of $+30^\circ$ \new{(see supplement Fig.~1)}.
There's no analogous effect for the first disc, but only because this
disc lies in the orbital plane; if it had a nonzero inclination, it
too would have $\omega_1^\mathrm{eff} \ne \omega_1$.  The effective
argument of pericentre may provide a better parameterization of
encounter geometry when comparing tidal responses of different
encounters.  As TT showed, tidal responses are typically strongest for
$\omega \sim 0^\circ$ and weakest for $\omega \sim \pm 90^\circ$.  In
the present study, since $\omega_2^\mathrm{eff} \sim 0^\circ$, most of
these discs should respond in a fairly uniform manner, with variations
in $\omega_2^\mathrm{eff}$ playing only a minor role.

\subsection{Orbit evolution}
\label{sec:orbevol}

\begin{figure}
\begin{center}
\epsfig{file=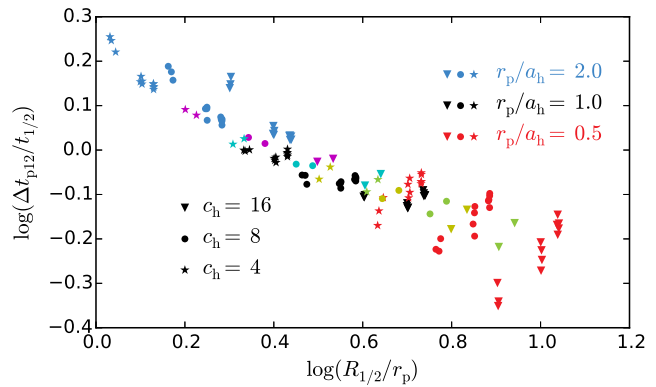,width=\columnwidth,
bbllx=78,bblly=258,bburx=489,bbury=503}
\caption{Scatter plot showing time between first and second passage,
normalized by the orbital period at the half-mass radius, versus the
interpenetration parameter.  As in Fig.~\ref{fig:interpen}, all $132$
encounters are plotted, symbol type indicates halo concentration, and
color indicates pericentric separation.
\label{fig:decaytime}}
\end{center}
\end{figure}

All of these galaxy pairs become bound \new{during} their first
passage, and subsequently fall back together.  The time between
\new{the} first and second \new{passages}, $\Delta t_\peri{12}$, is
comparable to the orbital period at the galactic half-mass radius,
$t_{1/2}$, since $t_{1/2}$ is the time-scale for a galaxy to rearrange
its mass distribution.  Fig.~\ref{fig:decaytime} plots $\Delta
t_\peri{12} / t_{1/2}$ against the interpenetration parameter $R_{1/2}
/ r_\peri{}$ (also see supplement Fig.~2).  Basically, all $132$
encounters have their second passage at a time $\Delta t_\peri{12}
\simeq (0.5 \mathrm{\,to\,} 2) \, t_{1/2}$ after their first passage.
Moreover, the variation in $\Delta t_\peri{12} / t_{1/2}$ is strongly
correlated with the degree of interpenetration, with the closest
encounters resulting in the most rapid orbit decay.

\begin{figure*}
\begin{center}
\epsfig{file=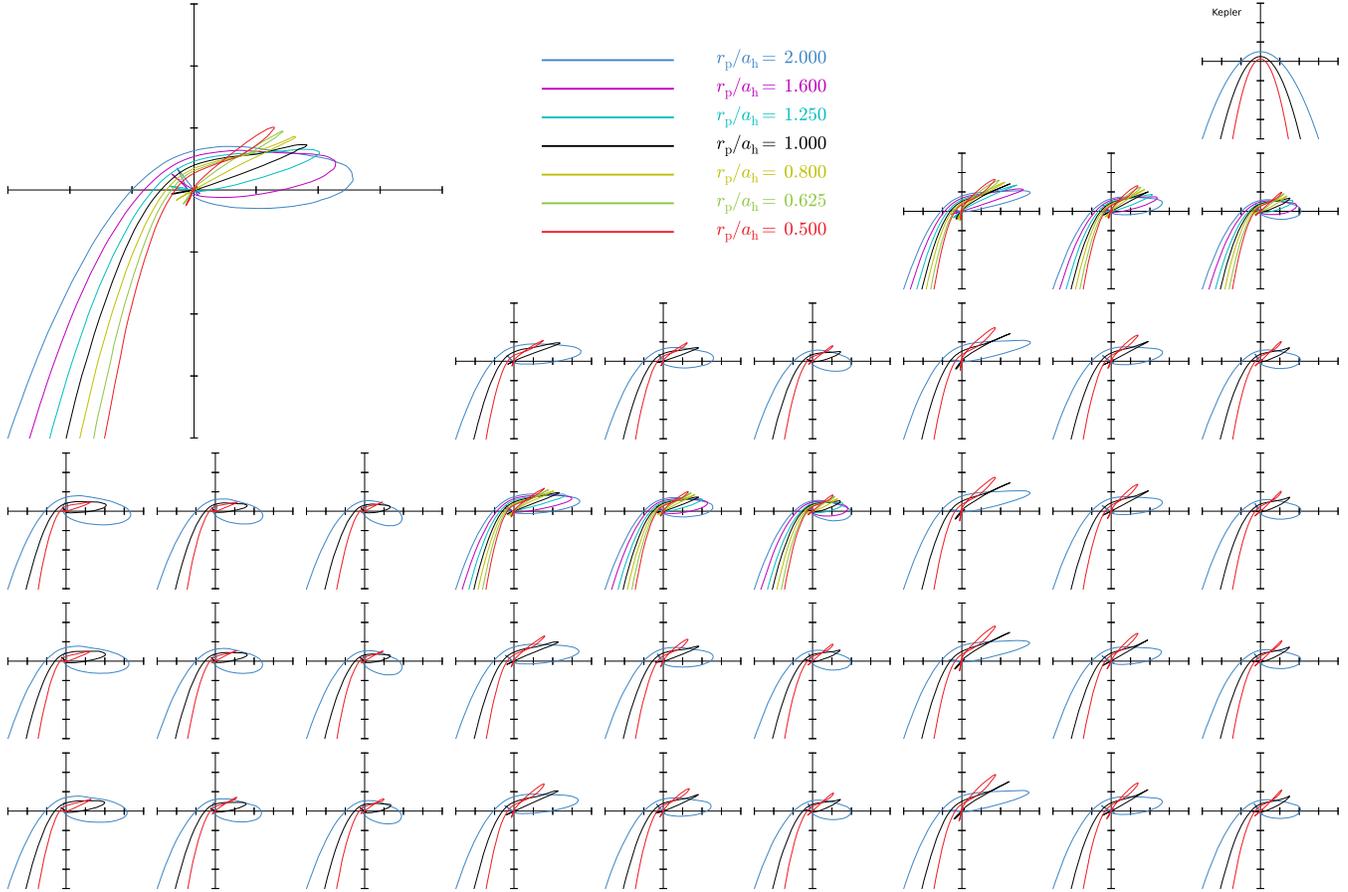,height=\textwidth,angle=270,
bbllx=68,bblly=39,bburx=544,bbury=753}
\caption{Relative orbital trajectories of all $132$ encounters,
\new{plotted on the orbital plane.  The encounters are} grouped into
$36$ ensembles with identical $f_\lumin$, $c_\halo$, and $\alpha_\disc
a_\halo$ values; line color indicates $r_\peri{} / a_\halo$.  The
layout of this \new{figure} matches Fig.~\ref{fig:vrot}; in the region
previously occupied by unstable models, the ensemble with $(f_\lumin,
c_\halo, \alpha_\disc a_\halo) = (0.1, 8, 3.0)$ is replotted on a
larger scale for better visibility; \new{fine details in other panels
may be best viewed electronically.}  The small plot labeled `Kepler'
shows representative parabolic trajectories. Tick marks are spaced $4
a_\halo$ apart.
\label{fig:trakxy}}
\end{center}
\end{figure*}

Fig.~\ref{fig:trakxy} presents relative orbital trajectories for the
entire sample of $132$ encounters, grouped into $36$ ensembles \new{--
one ensemble for each stable galaxy model}.  In these plots, the
position of galaxy~$2$ is shown with respect to galaxy~$1$.  All
encounters initially travel in a clockwise direction.  \new{After
first passage} the galaxies \new{are trapped} on bound orbits, which
typically attain apocentric separations of $\sim 12 a_\halo$ or less
for even the widest encounters (blue curves; $r_\peri{} / a_\halo =
2$).  Luminous fraction $f_\lumin$ and halo concentration $c_\halo$
both \new{systematically influence} orbital trajectory.  Decreasing
$f_\lumin$ shifts the position of the apocentre counter-clockwise,
while decreasing $c_\halo$ reduces the apocentric distance.  On the
other hand, disc compactness has almost no influence on these
trajectories; for the $f_\lumin = 0.2$ encounters, apocentric
separation decreases slightly as $\alpha_\disc a_\halo$ is reduced,
but no discernible effect is seen for smaller values of $f_\lumin$.
This is not very surprising since the disc is a small fraction of the
total mass.

Rather more surprising is the \textit{reversal} of orbital angular
momentum following close passages of extended, massive haloes.  This
can be seen, for example, in the large plot for ensemble $(f_\lumin,
c_\halo, \alpha_\disc a_\halo) = (0.1, 8, 3.0)$, where the oval curves
traced by the wider encounters give way to increasingly hairpin turns
at apocentre as $r_\peri{} / a_\halo$ is reduced.  For the two closest
members of the ensemble (green and red curves, for $r_\peri{} /
a_\halo = 0.625$ and $0.5$, respectively), the hairpin becomes a
self-crossing loop, and galaxy~$2$ falls back toward galaxy~$1$ on a
slightly \textit{counter-clockwise} path.  This curious behavior is
not limited to members of this ensemble; it appears uniformly in every
encounter with $f_\lumin \le 0.1$ and $r_\peri{} / a_\halo = 0.5$.

\begin{figure}
\begin{center}
\epsfig{file=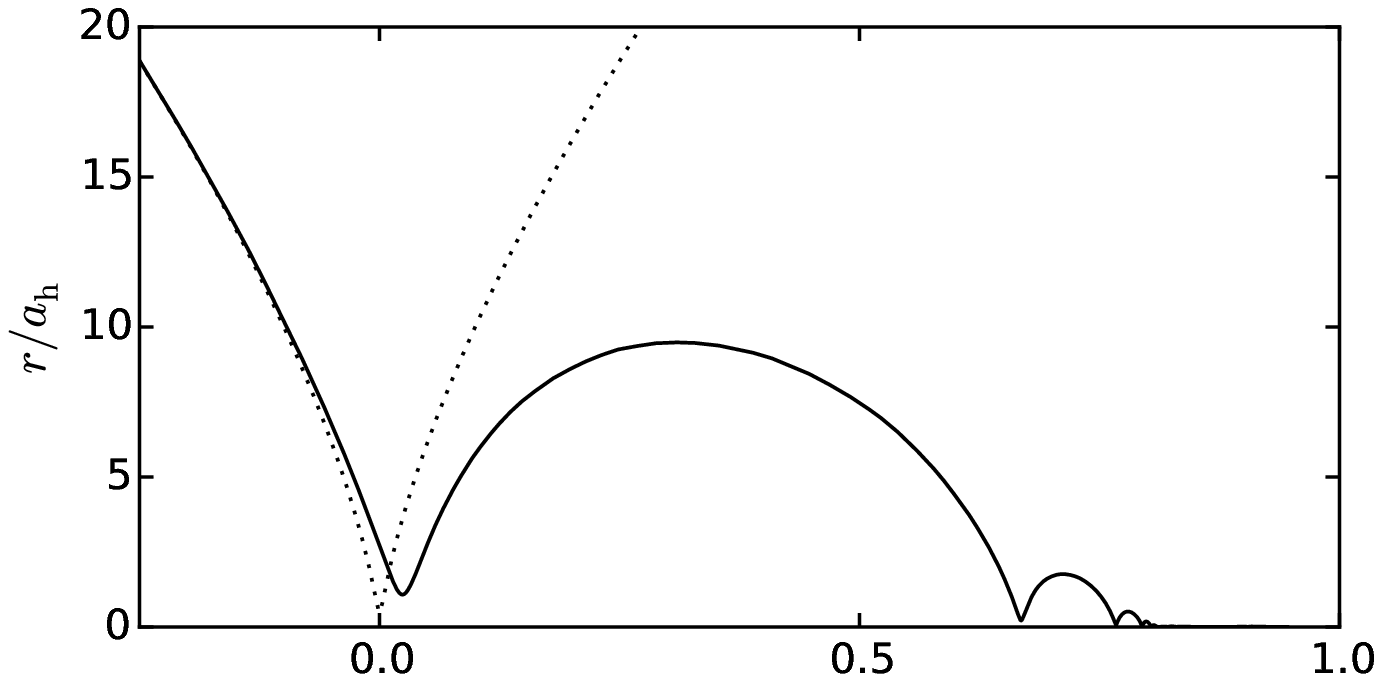,width=\columnwidth,
bbllx=82,bblly=272,bburx=489,bbury=488}
\epsfig{file=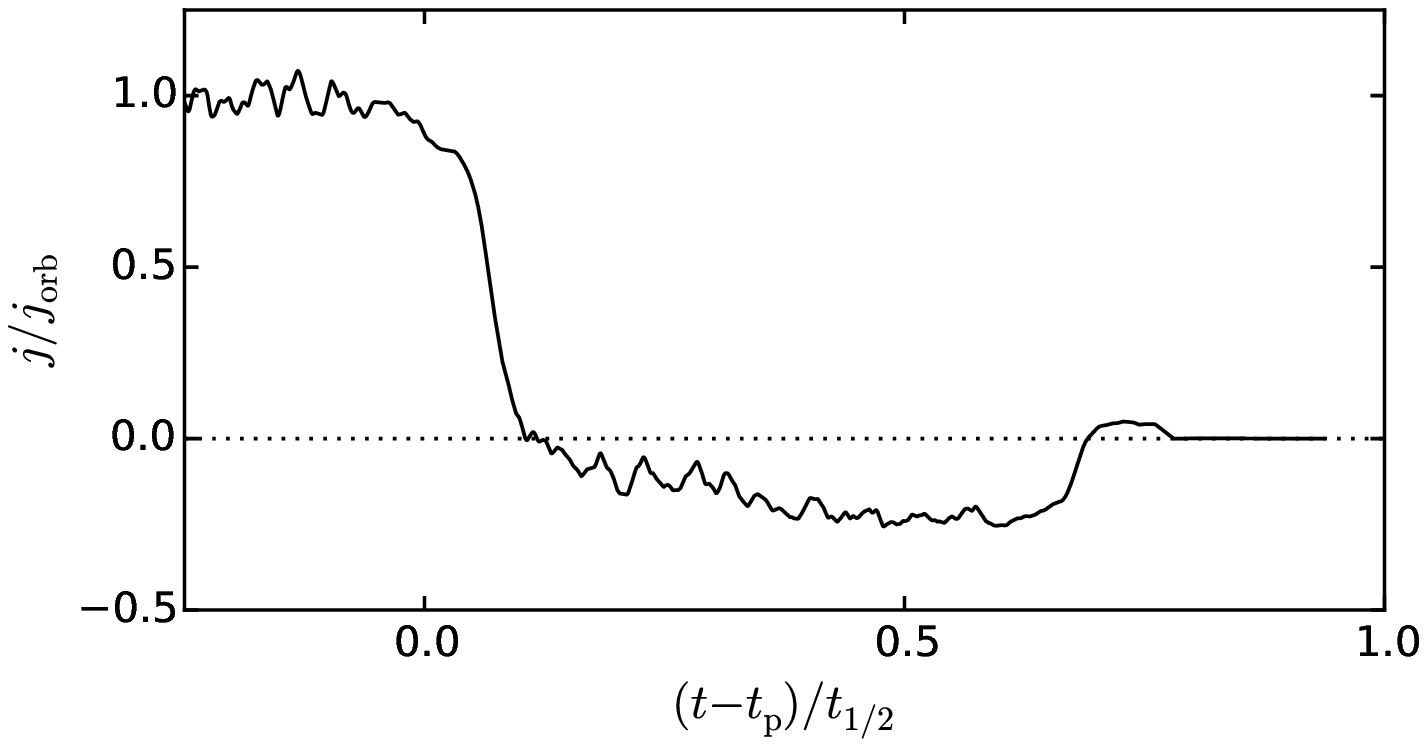,width=\columnwidth,
bbllx=82,bblly=272,bburx=489,bbury=488}
\caption{Evolution \new{of an encounter}, with parameters $(f_\lumin,
c_\halo, \alpha_\disc a_\halo, r_\peri{} / a_\halo) = (0.05, 16, 4.8,
0.5)$, \new{which exhibits reversal of orbital angular momentum}.  Top
panel shows separation as a function of time $t$, normalized \new{by}
$t_{1/2}$, the orbital period at the galactic half-mass radius. The
dotted curve is an equaivalent Keplerian orbit.  Bottom panel shows
\new{the specific} orbital angular momentum normalized to its initial
value; \new{note that $j$ changes sign after \new{both} the first and
second passages}.
\label{fig:orbrev}}
\end{center}
\end{figure}

Another view of this effect is provided by Fig.~\ref{fig:orbrev},
where the top and bottom panels show the separation between the galaxy
centres $r$ and their specific orbital angular momentum $j$,
respectively.  Prior to first passage, $j$ merely fluctuates about the
specific orbital angular momentum $j_\mathrm{orb}$; these fluctuations
are due to ongoing exchanges of linear momentum between each centre
and its own surrounding halo, magnified by the long lever arm afforded
by large values of $r$.  At first pericentre, $j$ has declined by only
$\sim 18$~percent, consistent with Fig.~\ref{fig:radvelperi}, but
shortly thereafter it drops dramatically as the tidally distorted
haloes rapidly absorb angular momentum.  This process continues well
past first pericentre, with $j$ finally changing sign when the galaxy
centres are $\sim 6 a_\halo$ apart and continuing to decrease until
somewhat after first apocentre.  At second pericentre the whole
process appears to repeat in a roughly self-similar pattern, with $j$
changing sign yet again shortly after $t_\peri{2}$.

\begin{figure}
\begin{center}
\epsfig{file=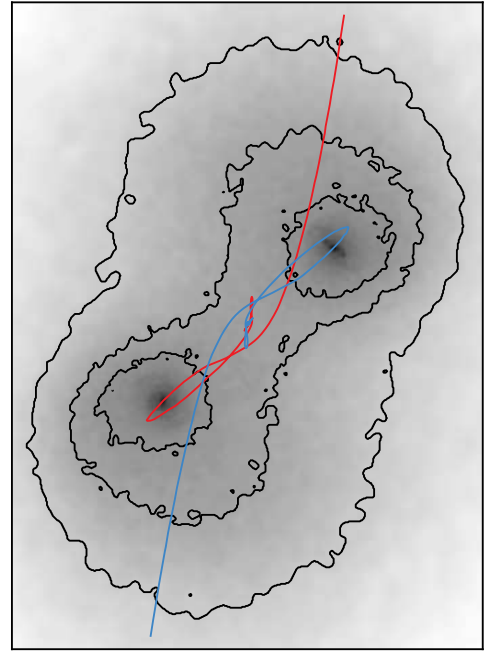,width=0.75\columnwidth,
bbllx=186,bblly=222,bburx=440,bbury=570}
\caption{Mass distribution of encounter in Fig.~\ref{fig:orbrev}
shortly before \new{first apocenter}, projected onto the orbital
plane.  Contours enclose $25$, $50$, and $75$ precent of the mass in
projection.  Red and blue curves are galaxy trajectories.
\label{fig:negtorque}}
\end{center}
\end{figure}

What accounts for this `extinction beyond the zero'\footnote{This
phrase has an interesting literary history, which the reader is
encouraged to discover.} of angular momentum?  \new{Any explanation
invoking dynamical friction will fail}; friction can reduce angular
momentum asymptotically to zero, but not beyond.  Instead, consider a
parabolic ($e = 1$), nearly head-on encounter of two extended, massive
haloes; after deeply interpenetrating, they will evolve toward a
prolate structure tumbling very slowly in the plane of their initial
orbits.  Now suppose these haloes each contain a self-gravitating
component (\new{traced, for example, by} a bulge) which, being much
smaller in radius, experiences the same encounter as hyperbolic and
grazing.  These bulges will be strongly deflected and will, for a
time, separate in a direction making a significant angle to the major
axis of the prolate structure formed by the two haloes.  As they do
so, they encounter a steep gravitational gradient which pushes them
back toward the major axis even before it halts their outward motion;
as a result, their orbital angular momenta reverse.
Fig.~\ref{fig:negtorque} shows the mass distribution of the encounter
in Fig.~\ref{fig:orbrev} just before first apocentre, at a time when
the specific angular momenta of the galaxies is rapidly decreasing;
the misalignment of the outer, roughly prolate bar and the inner
dumbbell will clearly torque the latter in a counter-clockwise
direction.

\begin{figure}
\begin{center}
\epsfig{file=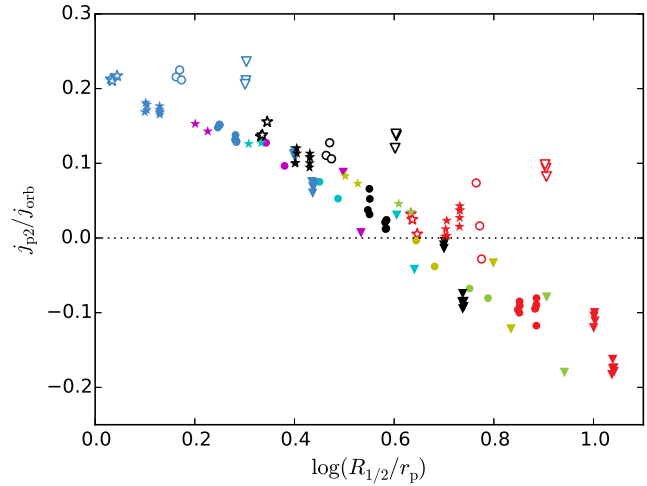,width=\columnwidth,angle=0,
bbllx=80,bblly=229,bburx=480,bbury=532}
\caption{\new{Relationship between interpenetration factor and orbital
angular momentum at second pericentre.  Symbol types and colors follow
Fig.~\ref{fig:interpen}; in addition, encounters with $f_\lumin = 0.2$
are plotted as open symbols, while those with $f_\lumin \le 0.1$ are
plotted as solid symbols.  Note that the $f_\lumin = 0.2$ encounters
are often outliers in what is otherwise a fairly tight relationship;
and do not, for any combination of parameters used here, exhibit
significant `extinction beyond the zero'.}
\label{fig:jzscale}}
\end{center}
\end{figure}

\new{Fig.~\ref{fig:jzscale} shows the relationship between orbital
angular momentum at \textit{second} pericentre, $j_\peri{2} /
j_\mathrm{orb}$, and the degree of interpenetration at first passage,
$R_{1/2} / r_\peri{}$.  All encounters with $j_\peri{2} /
j_\mathrm{orb} < 0$ have undergone extinction beyond zero.}  The most
striking examples involve deeply interpenetrating encounters between
extended ($c_\halo = 16$) and massive ($f_\lumin = 0.05$) haloes;
\new{this is entirely consistent with the scenario outlined above.  In
contrast, this form of orbit decay is almost never observed for
encounters with $f_\lumin = 0.2$ (open symbols in
Fig.~\ref{fig:jzscale}); typically, low-mass halos can't exert enough
gravitational torque to drive $j$ beyond the zero.}

Such violent orbit decay contradicts the assumption that the `final
encounters' of merging systems involve roughly circular orbits
\citep[e.g.][]{T+2000,RF2012}.  For a wide range of initial
conditions, the second passages of these \new{equal--mass} pairs are
nearly head-on and intensely disruptive, and a third passage and
merger follow very shortly thereafter.  \new{This point has not been
widely recognized.  \citet{TMvdVG2015} find a similar effect in a
simulated major merger, although in the case they present the orbital
angular momentum did not reverse until after the \textit{second}
passage.  It's unclear if encounters of unequal--mass pairs can also
evolve in this fashion.}

\section{DISC RESPONSE}

\begin{figure*}
\begin{center}
\vskip -1.5cm
\epsfig{file=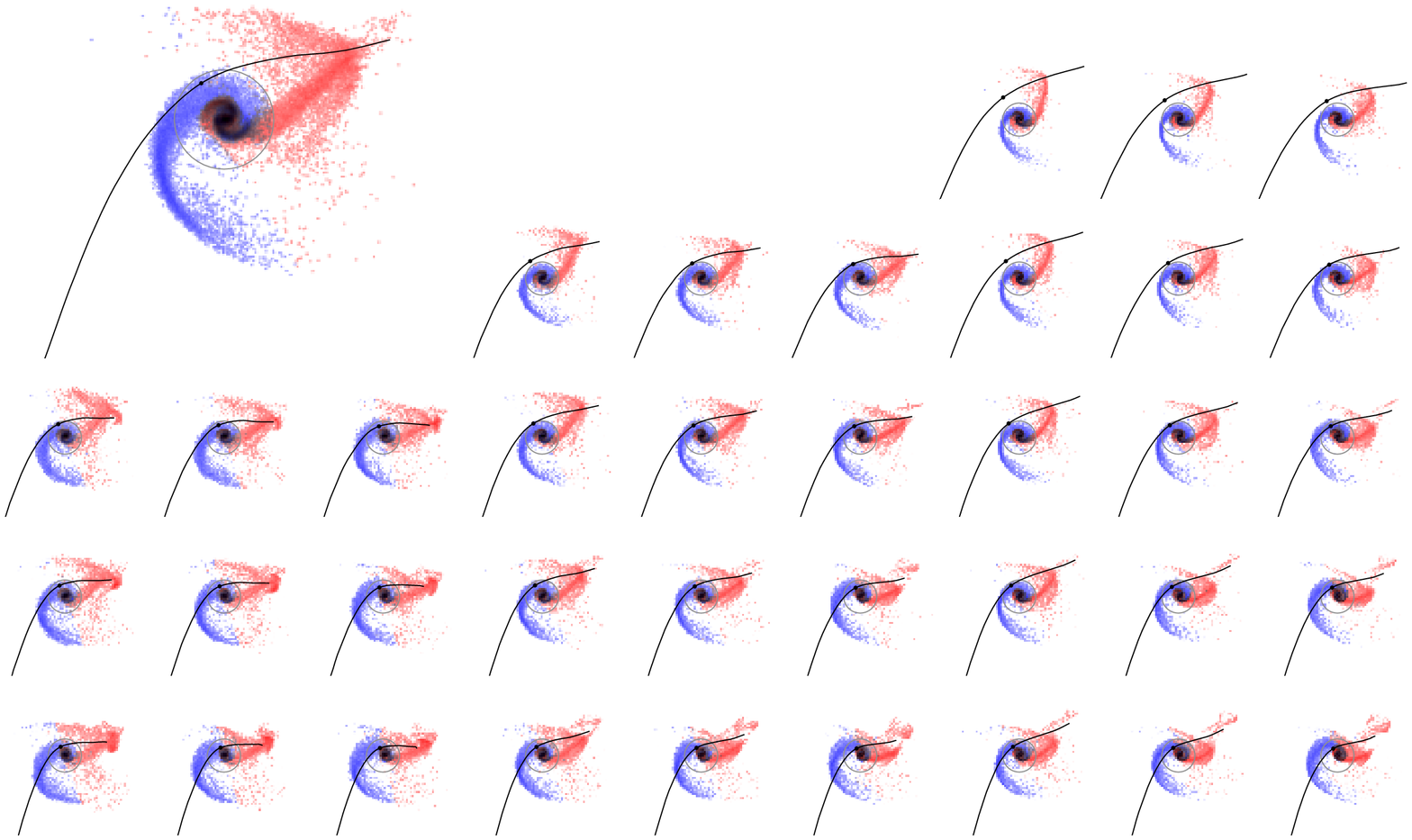,height=\textwidth,angle=270,
bbllx=68,bblly=39,bburx=544,bbury=753}
\epsfig{file=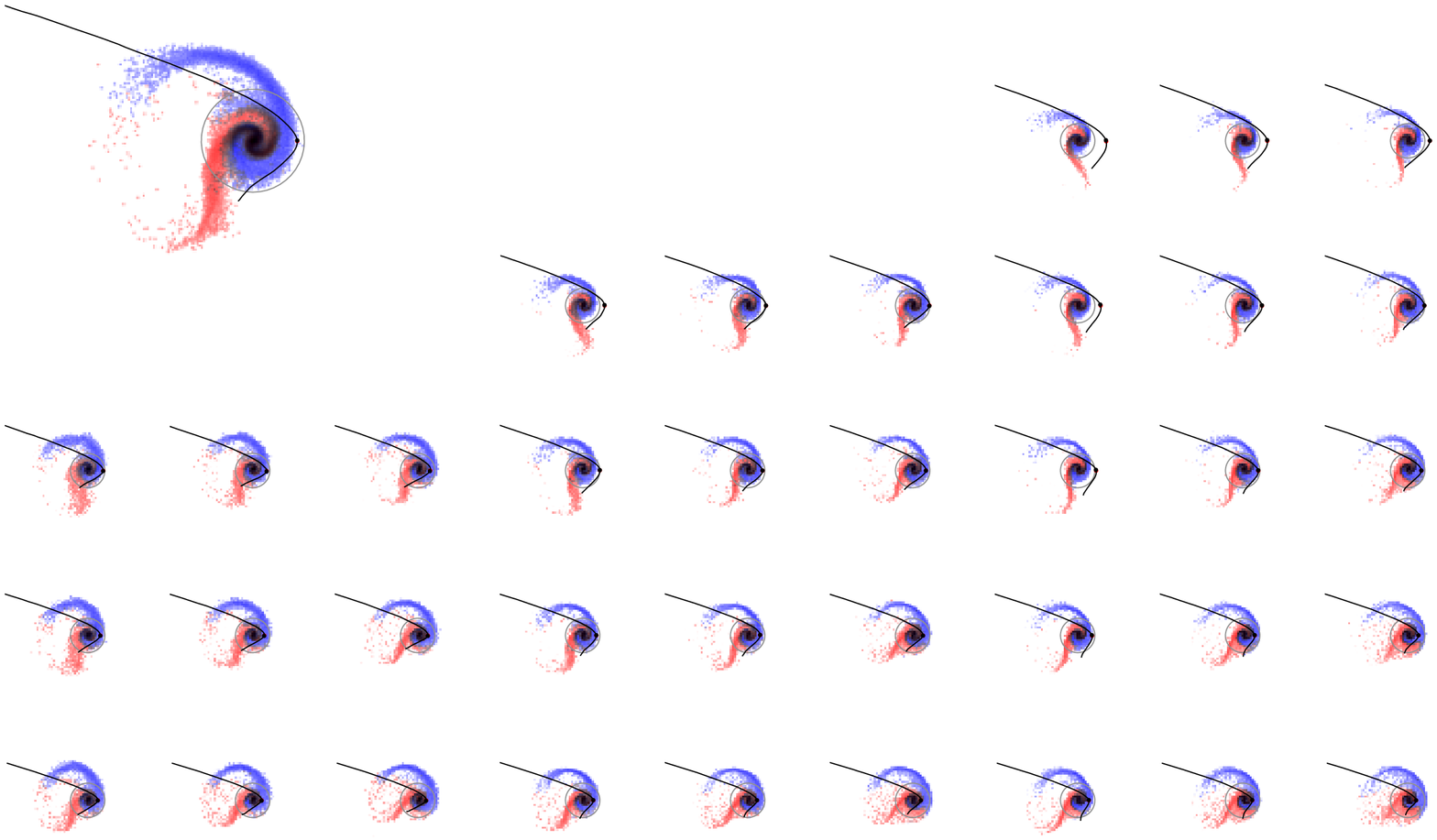,height=\textwidth,angle=270,
bbllx=68,bblly=39,bburx=544,bbury=753}
\caption{Disc response one rotation period after pericentre for
$r_\peri{}/a_\halo = 1$ encounters.  \new{Top and bottom} grids show
$i = 0^\circ$ and $i = 71^\circ$ discs, respectively, viewed face-on
\new{and scaled to keep $\alpha_\disc^{-1}$ constant}.  The layout of
each grid \new{mirrors Fig.~\ref{fig:trakxy}, with} the $(f_\lumin,
c_\halo, \alpha_\disc a_\halo, r_\peri{} / a_\halo) = (0.1, 8, 3.0,
1)$ \new{discs} replotted on a large\new{r} scale.  Black curves show
companion trajectores; a filled dot marks actual pericentre.  Grey
circles show the radius $5 \alpha_\disc^{-1}$.  Colors indicate
tidal classification; tails and bridges bodies are shown in blue and
red, respectively, \new{while regions containing both are rendered
black.  Note that colors are assigned to all bodies, but only those
beyond $5 \alpha_\disc^{-1}$ are counted as tidal features.}  See
supplement \new{Figs.~3 and~4} for other $r_\peri{}/a_\halo$ values.
\label{fig:discresp}}
\end{center}
\end{figure*}

\begin{figure*}
\begin{center}
\epsfig{file=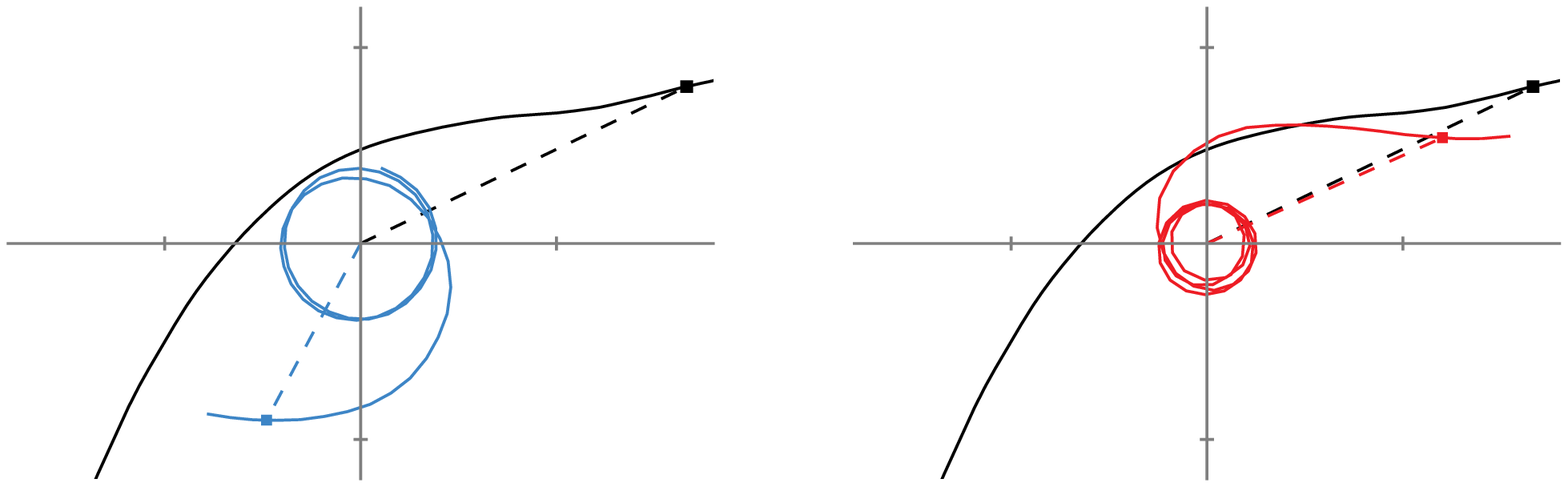,width=0.666\textwidth,
bbllx=35,bblly=311,bburx=594,bbury=481}
\caption{Classification of bodies in tidal features.  Each panel is
plotted with respect to the centre of the parent galaxy; \new{tick
marks are $10 \alpha_\disc^{-1}$ apart.}  The black curve shows the
trajectory of the companion galaxy, while the colored curve shows the
trajectory of the body being classified.  In each panel, filled
squares show positions when the body reaches its maximum radius for
the time interval \new{$t_\peri{1} < t < t_\mathrm{tid}$}, and the
angle between the dashed lines is $\psi_\apo{}$.  Left: tail
trajectory ($\cos \psi_\apo{} < 0$).  \new{Right}: bridge trajectory
($\cos \psi_\apo{} > 0$).
\label{fig:tdclexamp}}
\end{center}
\end{figure*}

\new{Fig.~\ref{fig:discresp} illustrates the relationship between
galaxy structure and tidal response.  At top right in the upper grid
are very compact discs situated deep within massive halos.  These
discs have been subject to in-plane yet fast and relatively distant
encounters and develop fairly symmetrical tidal features.  Moving down
these columns, disc size increases and the tidal response, while
stronger, becomes less symmetric, with bridges becoming noticeably
less coherent.  Moving to the left reduces the potential well depth
and encounter speed, both factors contributing to increased tidal
response.  In many of these slower cases, the bridge actually catches
up with and even `wraps around' the companion.}

\new{Inclined passages are presented in the bottom grid of
Fig.~\ref{fig:discresp}.  The discs in the upper right of this grid
again develop moderately symmetric tidal features.  Moving across this
grid a different morphology emerges, with the largest discs exhibiting
off-center ring-like features.  These rings result from roughly
perpendicular and deeply interpenetrating passages \citep{LT1976,
TS1977}; see Fig.~3 of \citet{B1992} for an illustration of this sort
of ring-making.}

\subsection{Identification of tidal features}

SW measured the strength of tidal features by counting all bodies at
distances $R > 10 \alpha_\disc^{-1}$ from their parent galaxy's
centre of mass.  They defined $T(t)$ to be the fraction of disc bodies
satisfying this criterion at time $t$, and took the maximum value,
$T_\mathrm{eff}$, as an effective measure of tidal response.  However,
this strategy has some limitations which became apparent in analyzing
the wide range of tidal encounters studied here.  First, while SW's
criterion is appropriate when focusing on \textit{long} tidal tails,
some discs exhibit definite tidal features which fit almost entirely
within a radius $R < 10 \alpha_\disc^{-1}$.  Second, the peak value,
while useful to show that elongated tidal features occur, does not
address the duration of their visibility.  Third, SW's criterion
counts tail \textit{and} bridge bodies indiscriminately.  In order to
make accurate statements about, e.g., the production of tidal tails in
different encounters, some method of sorting bodies into tidal
features is necessary.

On \new{a dynamical basis, tidal features should develop roughly} one
rotation period after pericentric passage.  Thus, rather than seeking
the instant when such features are maximized, each system is analyzed
at time $t_\mathrm{tid} = t_\peri{1} + t_\mathrm{rot}$, where
$t_\peri{1}$ is the actual time of first pericentre, and
$t_\mathrm{rot}$ is the rotation period at radius $R = 2 \,
\alpha_\disc^{-1}$.  This radius encloses $\sim 59$ percent of the
disc mass, so $t_\mathrm{rot}$ is close to the median rotation period
and provides a good overall measure.  Direct inspection of individual
discs at this time confirms that $R > 10 \alpha_\disc^{-1}$ is too
strict.  \new{A variety of} criteria based on some combination of each
body's initial radius, current radius, and maximum radius \new{were
tested}, \new{but in the end it proved most straightforward to count}
body $i$ as part of a tidal feature if
\begin{equation}
  R_i\new{(t_\mathrm{tid})} > 5 \alpha_\disc^{-1} \, ,
  \label{eq:tidal-def}
\end{equation}
\new{where $R_i(t) = |\vect{r}_i(t) - \overline{\vect{r}}_j(t)|$ is
the distance between body~$i$ and the centre of its parent
galaxy~$j$.}  The `optical radii' of disc galaxies typically extend to
$\sim 5 \alpha_\disc^{-1}$, so \new{(\ref{eq:tidal-def}) basically
identifies tidal features as material} beyond the optical radius.
\new{A} light grey circle superimposed on each image \new{in
Fig.~\ref{fig:discresp}} shows the radius \new{$R = 5
\alpha_\disc^{-1}$}.  In some cases, bridges and tails continue inward
to smaller radii, while in others the discs themselves appear to
extend slightly beyond, but on the whole \new{this} seems to be a
reasonable working definition of tidal material.

\begin{figure*}
\begin{center}
\epsfig{file=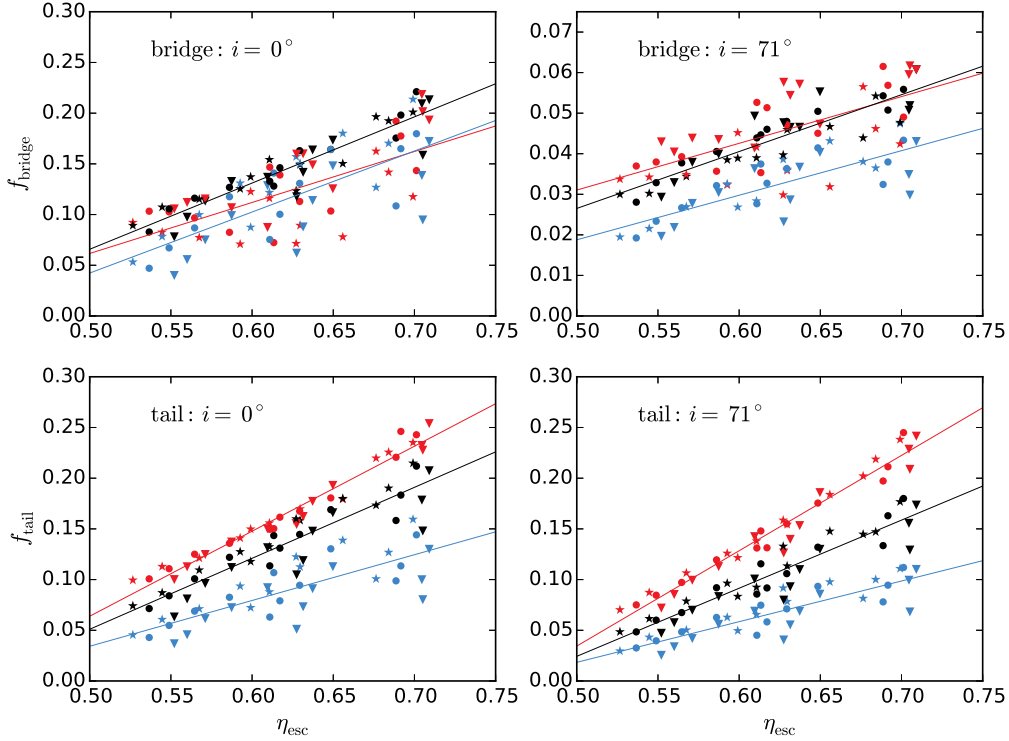,height=0.75\textwidth,angle=270,
bbllx=96,bblly=73,bburx=544,bbury=682}
\caption{Scatter plots \new{showing amount} of tidal material versus
$\new{\eta_\mathrm{esc} =} \sqrt{\new{2} / \mathcal{E}}$ \new{for the
primary sample of $108$ encounters}, broken down by feature
\new{classification} and disc \new{inclination}.  Here,
$f_\mathrm{bridge}$ and $f_\tail{}$ are the fractions of disc bodies
classified as bridges and tails, \new{measured one rotation period
after first pericentre}.  \new{As in Fig.~\ref{fig:interpen},} color
indicates $r_\peri{} / a_\halo$, \new{while symbol type indicates
$c_\halo$}.  Solid lines are linear fits for different $r_\peri{} /
a_\halo$ values.  Note that the panel for the $i = 71^\circ$ bridge
(top right) has a $y$-axis range \new{one-quarter the range of} the
other three \new{panels}.
\label{fig:tidalfrac}}
\end{center}
\end{figure*}

Having identified the bodies belonging to tidal features, the next
step is to classify them as members of bridges or tails.
Fig.~\ref{fig:tdclexamp} illustrates the classification algorithm,
which \new{works by analyzing individual trajectories}.  Follow
\new{each} body \new{from time $t_\peri{1}$} until time
$t_\mathrm{tid}$, and let $t_\apo{}$ be the instant when the body's
distance from its parent is greatest (in many but not all cases,
$t_\apo{} = t_\mathrm{tid}$).  At \new{time} $t_\apo{}$, construct
unit vectors $\widehat{\vect{n}}_\apo{}$ and
$\widehat{\vect{q}}_\apo{}$ from the parent to the companion and the
body, respectively, and let $\cos \psi_\apo{} =
\widehat{\vect{n}}_\apo{} \cdot \widehat{\vect{q}}_\apo{}$.
\new{Bodies which are on the side opposite the companion at $t_\apo{}$
have $\cos \psi_\apo{} < 0$ and are classified as tail particles,
while those on the same side have $\cos \psi_\apo{} > 0$ and are
classified as bridge particles.  Fig.~\ref{fig:discresp} uses colors
to indicate tidal classifications, with tails in blue and bridges in
red.}

Fig.~\ref{fig:tidalfrac} plots fractions of disc bodies in tidal
features, organized by disc inclination and feature classification.
Instead of using $\mathcal{E}$ as defined by
(\ref{eq:sw-escape-param}), the horizontal axis is
$\new{\eta_\mathrm{esc} =} \sqrt{\new{2} / \mathcal{E}} = \sqrt{2} \,
v_\mathrm{c} / v_\mathrm{e}$.  \new{Up to a constant factor,
$\eta_\mathrm{esc}$ is just the ratio of circular to escape velocity;
following SW, this ratio is evaluated at radius $R = 2 \,
\alpha_\disc^{-1}$.}  This parameterization is useful because it
produces roughly linear trends with $f_\tail{}$; moreover, \new{the
normalization insures that} a Keplerian potential yields
\new{$\eta_\mathrm{esc} = 1$, which serves as a convenient reference.
Note that SW's criterion $\mathcal{E} \lesssim 6.5$ is equivalent to
$\eta_\mathrm{esc} \gtrsim 0.55$.}

\new{In Fig.~\ref{fig:tidalfrac}}, symbol color indicates pericentric
separation $r_\peri{} / a_\halo$, and the solid lines are linear fits
for $r_\peri{} / a_\halo = \new{0.5}$ (red), $\new{1.0}$ (black), and
$\new{2.0}$ (blue).  These plots reveal some interesting
relationships.  As SW found, the parameter $\mathcal{E}$ (or
equivalently, $\new{\eta_\mathrm{esc}}$) is strongly correlated with
tidal fraction (see also supplement Fig.~5).  This correlation is
particularly \new{striking} for tidal tails, while for bridges the
scatter is considerably larger.  For tails, the second parameter which
determines tidal fraction is pericentric separation; $f_\tail{}$
increases monotonically as $r_\peri{} / a_\halo$ decreases.  This
makes sense, since closer encounters produce stronger tides; indeed,
this plot suggests that still closer encounters may yield even larger
tail fractions.  Inclination $i$ enters as a third parameter,
influencing the slope of the relationship between
$\new{\eta_\mathrm{esc}}$ and $f_\tail{}$.

\begin{figure*}
\begin{center}
\begin{tabular}{lr}
\epsfig{file=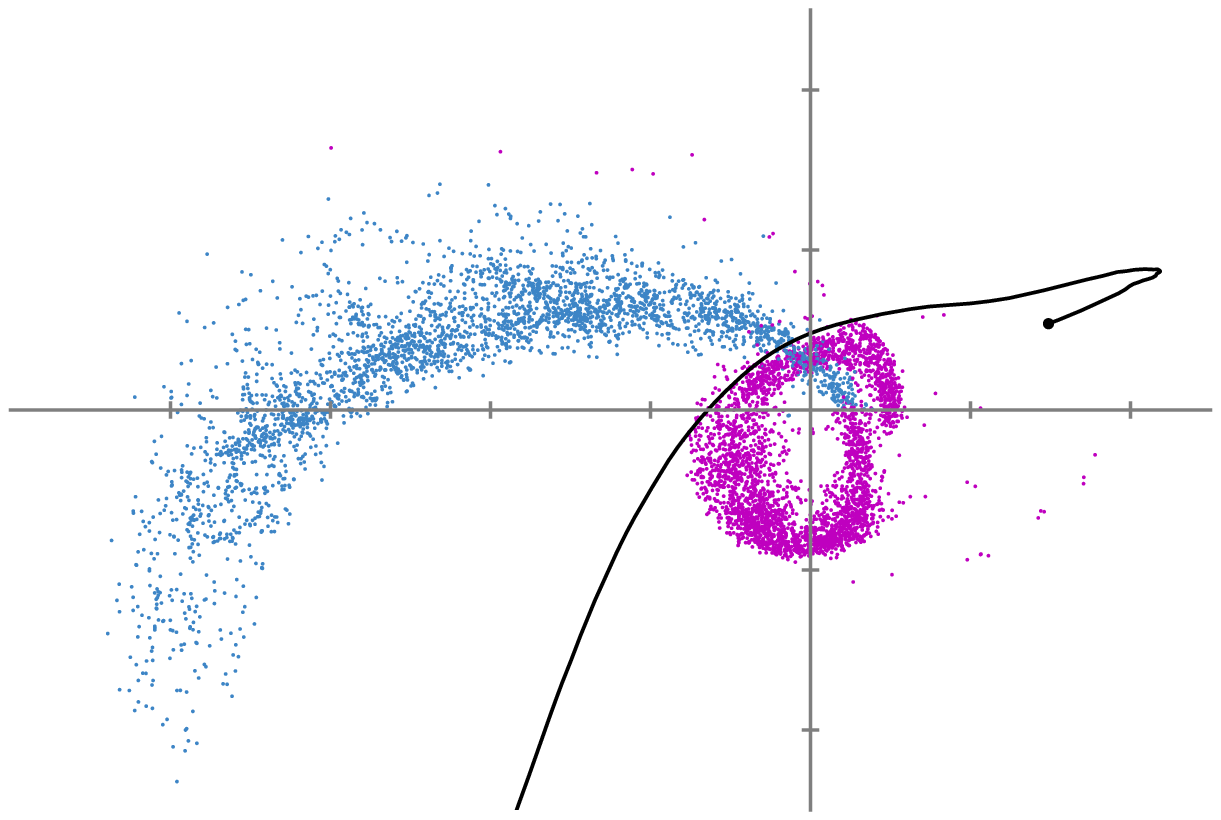,width=\columnwidth,
bbllx=132,bblly=280,bburx=539,bbury=512}&%
\epsfig{file=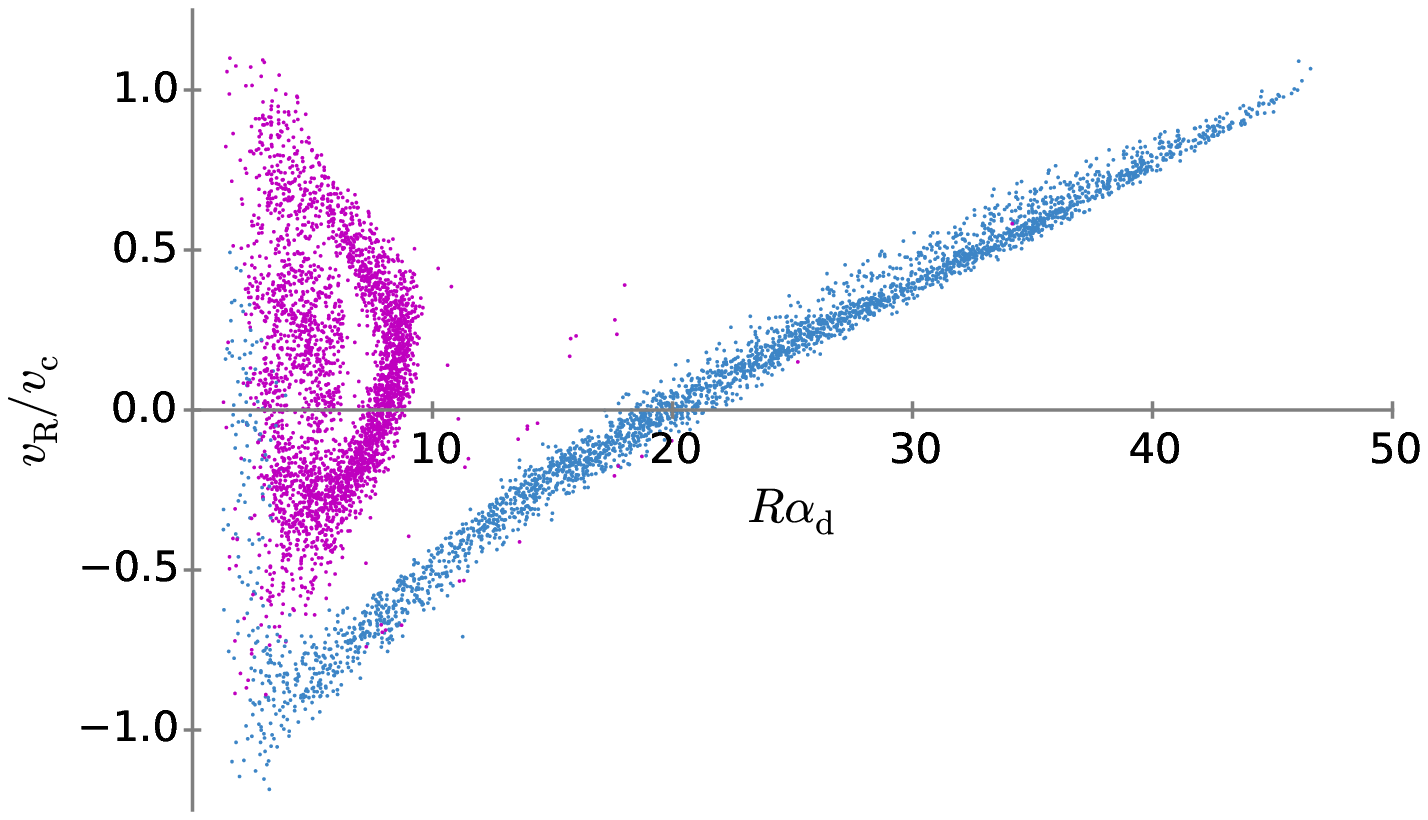,width=\columnwidth,
bbllx=80,bblly=280,bburx=487,bbury=512}\\
\end{tabular}
\caption{Reaccretion of the tidal tail from an $i = 0$ disc.  All the
bodies plotted were classified as tail material (blue) at time
$t_\mathrm{tid}$, \new{but by} the time shown, between first apocentre
and second pericentre, roughly half have been reclassified as loop
material (magenta).  Left: projection onto the disc plane, with the
parent galaxy at the origin; tick marks are $10 \alpha_\disc^{-1}$
apart.  The black curve shows the companion's relative trajectory.
Right: radius $R$ versus radial velocity $v_\mathrm{R}$, scaled by
disc scale length $\alpha_\disc^{-1}$ and circular velocity
$v_\mathrm{c}$ at $2 \alpha_\disc^{-1}$, respectively.  \new{All of
these tail bodies remain bound to the merger remnant.}
\label{fig:tailreturn}}
\end{center}
\end{figure*}

For bridges, the situation is more complex.  The tidal fraction
increases as $r_\peri{} / a_\halo$ is reduced from \new{$2.0$} to
\new{$1.0$}, but further reduction has the opposite effect, and the
scatter about the linear fits becomes larger, especially for the $i =
0$ disc.  It appears that these closer passages are so deeply
interpenetrating that bridge formation is suppressed, and this
suppression is especially effective for in-plane encounters.

The $i = 0^\circ$ discs typically develop comparable bridge and tail
fractions, although tails are somewhat favored as $r_\peri{} /
a_\halo$ decreases.  On the other hand, virtually \textit{all} of the
\new{tidal features from the $i = 71^\circ$ discs are tail-dominated},
often by \new{factors} of $\sim 3$ or more.  This result harks back to
TT's fig.~15, which shows the bridge bodies dwindling relative to
tails as inclination increases.  It's not clear why bridges are more
sensitive to inclination than tails; the quasi-resonant formalism of
\citet{DOVFGH2010} may be applicable \new{to this question, but
perturbation expansions} to a fairly high order \new{appear needed to
address it}.

\subsection{Lifetimes of tidal tails}
\label{sec:lifetail}

All of the galaxy models examined in this paper can produce fairly
substantial bridges and tails, especially \new{when involved in close}
encounters.  However, these features don't always persist.  In
equal-mass encounters, little \new{or no} tidal material reaches
escape velocity, so tails and bridges are destined to fall back into
their parent galaxies, creating complex systems of reaccreted tidal
loops as shown in Fig.~\ref{fig:tailreturn}.

\begin{figure*}
\begin{center}
\epsfig{file=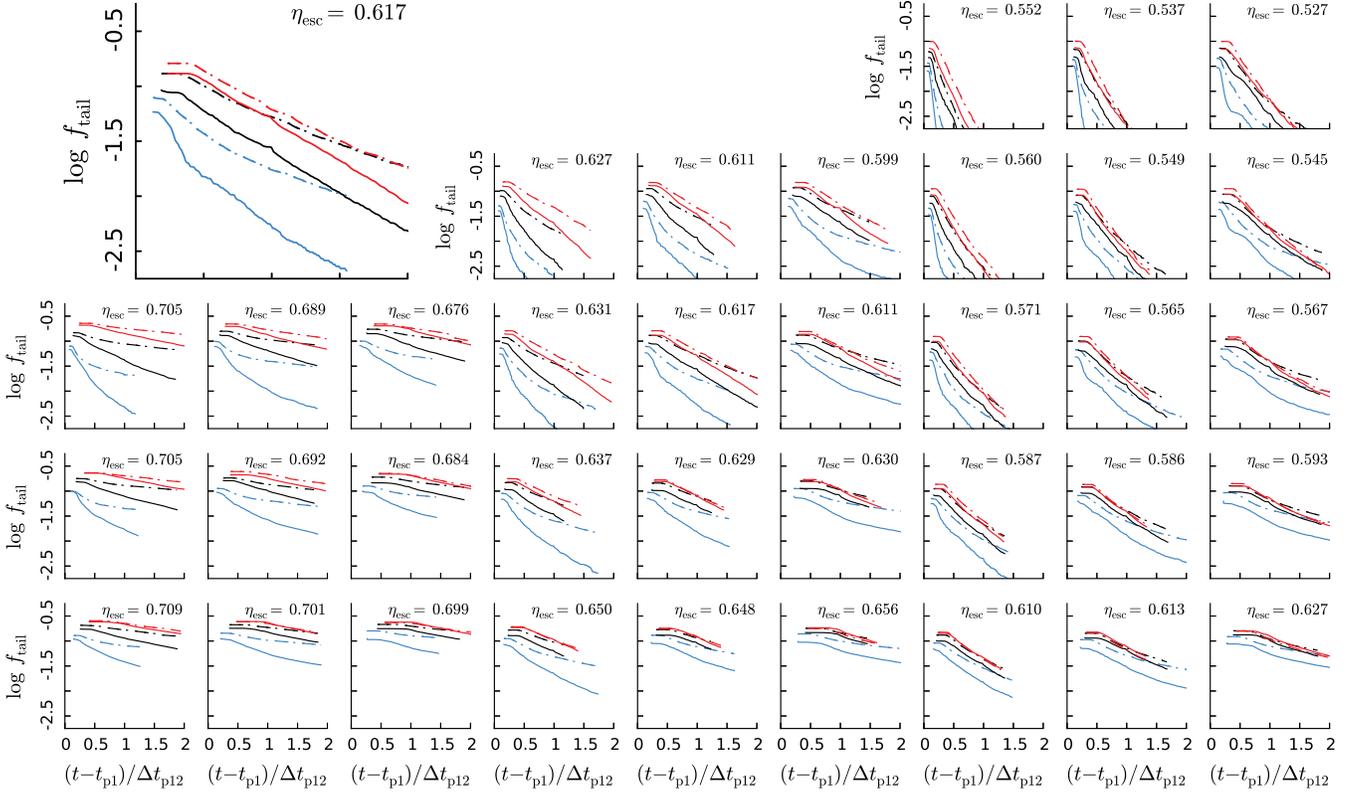,height=\textwidth,angle=270,
bbllx=146,bblly=102,bburx=491,bbury=684}
\caption{Evolution of tail fractions for the primary encounter sample.
The layout \new{follows Fig.~\ref{fig:trakxy}; each panel represents
an ensemble of encounters with different $r_\peri{}/a_\halo$ values,
indicated by line color, and the $(f_\lumin, c_\halo, \alpha_\disc
a_\halo) = (0.1, 8, 3.0)$ ensemble is replotted on a larger scale.}
Each panel plots log tail fraction against time since first
pericentre, measured in units of the time between first and second
passages, \new{and the value of $\eta_\mathrm{esc}$ for the galaxy
model used in each panel appears at the top}.  Dot-dashed and solid
lines show \new{tail fractions for} $i = 0^\circ$ and $i = 71^\circ$
discs, respectively.  See supplement Figs.~6 and~7 for further
details on tail evolution. 
\label{fig:tailevol}}
\end{center}
\end{figure*}

This process is easier to analyze for tidal tails, which, \new{once
they are launched, evolve} mostly under the influence of their parent
galaxy, with relatively little ongoing interference from the
companion.  The basic \new{technique} is to track each tail body until
its next encounter with its parent, and at that time reclassify it as
belonging to a tidal loop instead of a tail.  Naively, this can be
done by finding the next local minimum of $R_i(t)$, the distance
between tail body~$i$ and the centre of its parent galaxy~$j$.
Numerical noise in $\overline{\vect{r}}_j$ may trigger
reclassification prematurely when body $i$ is near apocentre; to avoid
this difficulty, the minimum \new{is required to} satisfy $R_i(t) \le
\new{P} R_i^\mathrm{max}$, where $\new{P} = 0.95$ and
$R_i^\mathrm{max} = \max(R_i(t))$ is the tail body's maximum distance
from its parent.  This simple method works well until shortly before
the system's second pericentre, but fails when the acceleration of the
parent galaxy causes rapid changes in $R_i(t)$.  A better-behaved
function $R^\prime_i(t)$ can be defined by smoothly interpolating
between $\overline{\vect{r}}_j$ and the system centre-of-mass position
$\overline{\vect{r}}_\mathrm{cm} = (\overline{\vect{r}}_1 +
\overline{\vect{r}}_2) / 2$ as second pericentre approaches:
\begin{equation}
  R^\prime_i(t) =
    |\vect{r}_i - (\new{Q} \overline{\vect{r}}_j +
                   (1 - \new{Q}) \overline{\vect{r}}_\mathrm{cm})| \, ,
\end{equation}
where $\new{Q = \min(1, d_\mathrm{min} \alpha_\disc / 5)}$ depends on
the \textit{minimum} distance $d_\mathrm{min}$ between the two galaxy
centres up to time $t$.  In addition, the criterion $R^\prime_i(t) \le
\new{P} R_i^\mathrm{max}$ can be tightened as the galaxies approach
each other by setting $\new{P = 0.6 + 0.35 \, \exp(- (d_\mathrm{min}
\alpha_\disc / 5)^{-4})}$.  With these adjustments, tail reaccretion
can be followed through multiple pericentric passages and ultimately
merger.

\begin{figure*}
\begin{center}
\epsfig{file=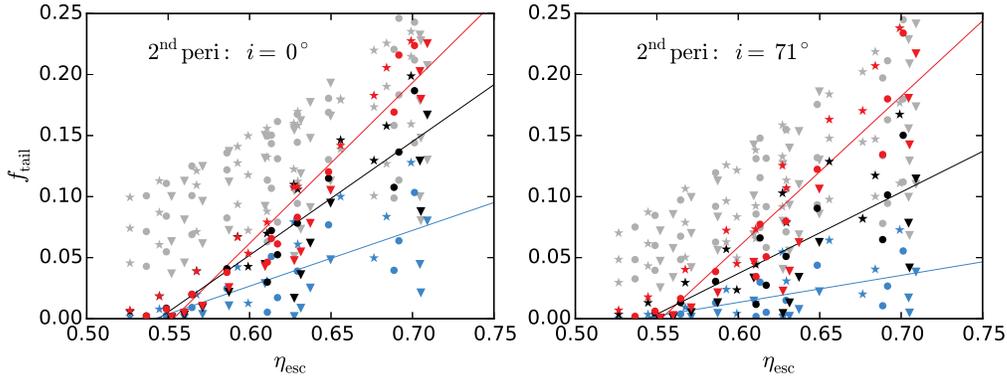,height=0.75\textwidth,angle=270,
bbllx=96,bblly=75,bburx=321,bbury=682}
\caption{Scatter plots of tail fraction at second pericentre, $t =
t_\mathrm{p2}$.  \new{As in Fig.~\ref{fig:tidalfrac}}, color and
symbol type indicate $r_\peri{} / a_\halo$ and \new{$c_\halo$, and the
solid lines are linear fits for each $r_\peri{} / a_\halo$}.  The
light grey symbols show tail fractions at time $t = t_\mathrm{tid}$,
one rotation period after first pericentre.
\label{fig:tailfrac}}
\end{center}
\end{figure*}

Fig.~\ref{fig:tailevol} shows how tail fractions $f_\tail{}$ evolve
with time.  At time $t_\mathrm{tid}$, when tidal features are first
identified, bodies at the bases of the tails are typically near
apocentre; consequently, tail fractions are initially almost constant.
However, reaccretion commences as these bodies fall back toward their
parents, and the tail fraction decreases monotonically thereafter.
Meanwhile, the galaxies themselves, after loitering near first
apocentre, \new{fall} back toward each other.  By second pericentre,
tail fractions have \new{often decreased} quite dramatically.

One might expect that closer encounters, which yield more tidal
material \textit{and} decay faster, would maximize tail fractions at
later times.  This is \new{confirmed} by Fig.~\ref{fig:tailevol};
\new{in each panel, the $r_\peri{} / a_\halo = 0.5$ curves (red) start
higher and usually decline more gradually than their $1.0$ (black) and
$2.0$ (blue) counterparts.  Another trend evident within individual
panels involves disc inclination; the $i = 71^\circ$ (solid) tails
often decline more steeply than the corresponding $i = 0^\circ$
(dot-dashed) tails.}

\new{Comparison between panels in Fig.~\ref{fig:tailevol} shows that
reaccretion rates, indicated by the slopes of the various curves, are
generally anticorrelated with $\eta_\mathrm{esc}$.  Galaxies with
$\eta_\mathrm{esc} \lesssim 0.55$, plotted at top of the right-hand
three columns of this figure, reaccrete so rapidly that they arrive at
second passage with no visible tails to speak of.  Further down these
columns, $\eta_\mathrm{esc}$ increases and the reaccretion rate
diminishes.  This general trend continues across the rest of the
figure, and galaxies with $\eta_\mathrm{esc} \gtrsim 0.70$, found in
the left-hand columns, typically reaccrete their tails very slowly,
and often reach second passage still festooned with massive tidal
tails.  Note, however, that the relationship between reaccretion rate
and $\eta_\mathrm{esc}$ has some scatter; within each group of three
columns, representing different halo concentrations for a given
luminous fraction, the rate of reaccretion decreases as $c_\halo$ is
reduced.  These trends are corroborated by supplement Fig.~8, which
shows the effects of $\eta_\mathrm{esc}$ and $c_\halo$ explicitly.}

The amount of tail material visible at second pericentre is further
examined in Fig.~\ref{fig:tailfrac}, which reproduces the layout of
the lower panels in Fig.~\ref{fig:tidalfrac}.  Since these encounters
merge shortly after second passage (e.g., by $(t - t_\peri{1}) /
\Delta t_\peri{12} = 1.25$), this plot provides an upper limit to the
amount of tail material merger remnants are likely to display.  The
roughly linear relationships between $f_\tail{}$ and
$\new{\eta_\mathrm{esc}}$ noted in Fig.~\ref{fig:tidalfrac} become
steeper as a result of tail reaccretion, \new{and halo concentration
emerges as an additional parameter; both of these effects follow
naturally from the trends seen in Fig.~\ref{fig:tailevol} and
supplement Fig.~8. Merger remnants with} conspicuous tails (e.g.,
$f_\tail{} \gtrsim 0.05$) appear to require \new{$\eta_\mathrm{esc}
\gtrsim 0.60$ (i.e., $\mathcal{E} \lesssim 5.5$)}; while many of the
models with lower-mass haloes satisfy this condition, only those
$f_\lumin = 0.05$ models which have relatively extended discs can do
so.  \new{Moreover, distant encounters frequently fail to produce
remnants with $f_\tail{} \gtrsim 0.05$ even when $\eta_\mathrm{esc}
\gtrsim 0.60$.}

\section{TIDAL CONFIGURATIONS}

\new{As shown in the previous section,} the strength of the tidal
response provides information about progenitor structure, but \new{the
the overall \textit{configuration} of the tidal response -- in other
words, the strength \textit{and} the morphology, taken together --}
may yield further constraints.  SW note that such `constraints are
potentially very powerful if dynamical modeling is combined with
detailed observation', but this idea has not yet been tested.  As a
first step toward this goal, we can ask \new{if} the relationship
between progenitor structure and tidal configuration exhibits simple
patterns which might be used to deduce halo properties?

One approach to this question is to systematically compare tidal
configurations produced in encounters with different progenitor
structures.  A pair of encounters which consistently mimic each
other's tidal configurations \new{will be impossible to distinguish}
observationally; such pairs are `degenerate'.  In the present tests,
the viewing direction and encounter geometry will be fixed \textit{a
priori}.  This is a fairly drastic simplification; in practice,
detailed modeling of interacting systems attempts to infer these
geometric parameters from the observed morphology and kinematics
\citep[e.g.,][]{BH2009}.  However, in some rather limited
circumstances the encounter and viewing geometry can be determined
independently of other parameters (for example, an encounter between
two discs with inclinations $i_1 = i_2 = 0$, viewed face-on to the
orbital plane, can be recognized as such from line-of-sight velocity
data).  The time since first passage, on the other hand, must be taken
as an unknown.

\subsection{Comparison procedure}
\label{sec:cmp_proc}

\begin{figure*}
\begin{center}
\begin{tabular}{ccc}
\epsfig{file=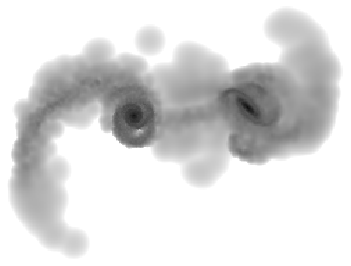,width=2.0in}&%
\epsfig{file=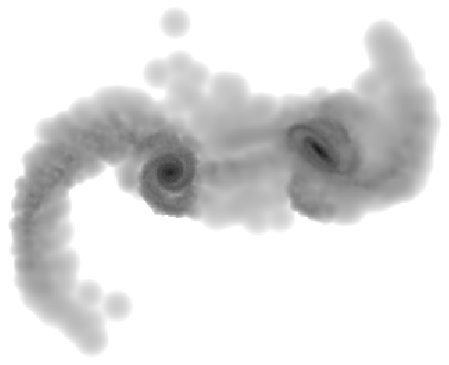,width=2.0in}&%
\epsfig{file=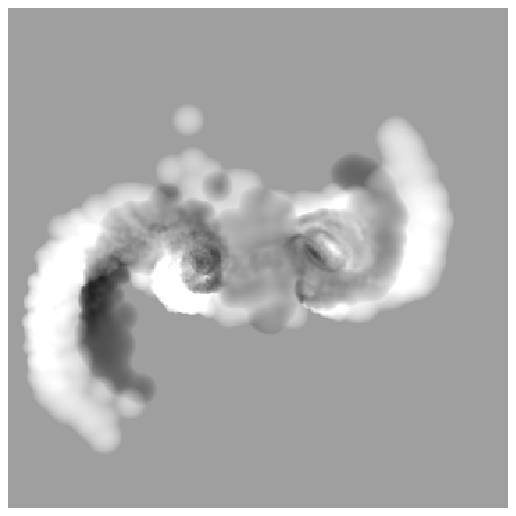,width=2.0in}\\
\epsfig{file=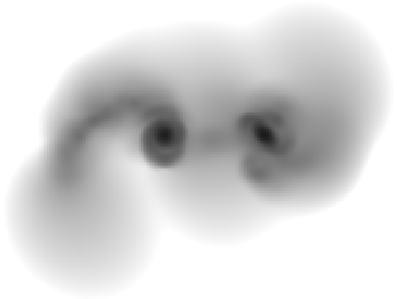,width=2.0in}&%
\epsfig{file=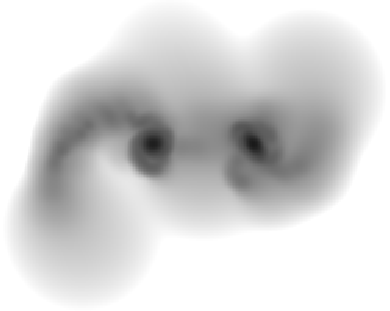,width=2.0in}&%
\epsfig{file=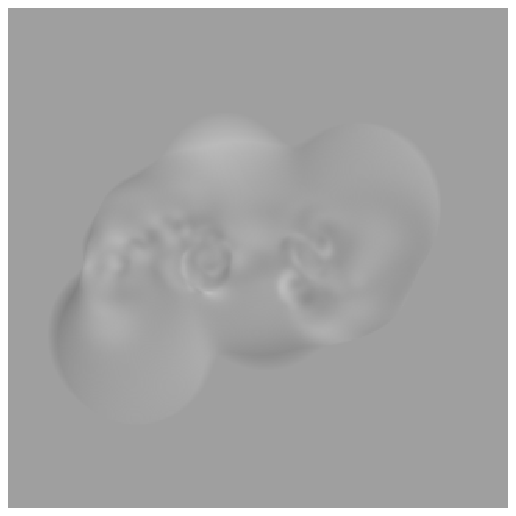,width=2.0in}\\
\end{tabular}
\caption{Two different \new{encounters which produce} similar
configurations. Left: \new{reference encounter} with parameters
$(f_\lumin, c_\halo, \alpha_\disc a_\halo, r_\peri{} / a_\halo) =
(0.1, 8, 3, 2)$, shown at \new{time $t_\mathrm{ref} = t_\apo{1}$}.
Middle: \new{comparison encounter} with parameters $(0.1, 16, 2.4,
2)$, shown \new{at the time yielding the best match to the reference
encounter}. Right: \new{difference between reference and comparison
images}.  Top: original simulation data, yielding ${D} = 0.395$.
Bottom: data transformed to place galaxy centres at $(\pm 1, 0, 0)$
and further smoothed, yielding ${D} = 0.049$.
\label{fig:imgmatch}}
\end{center}
\end{figure*}

\begin{figure}
\begin{center}
\epsfig{file=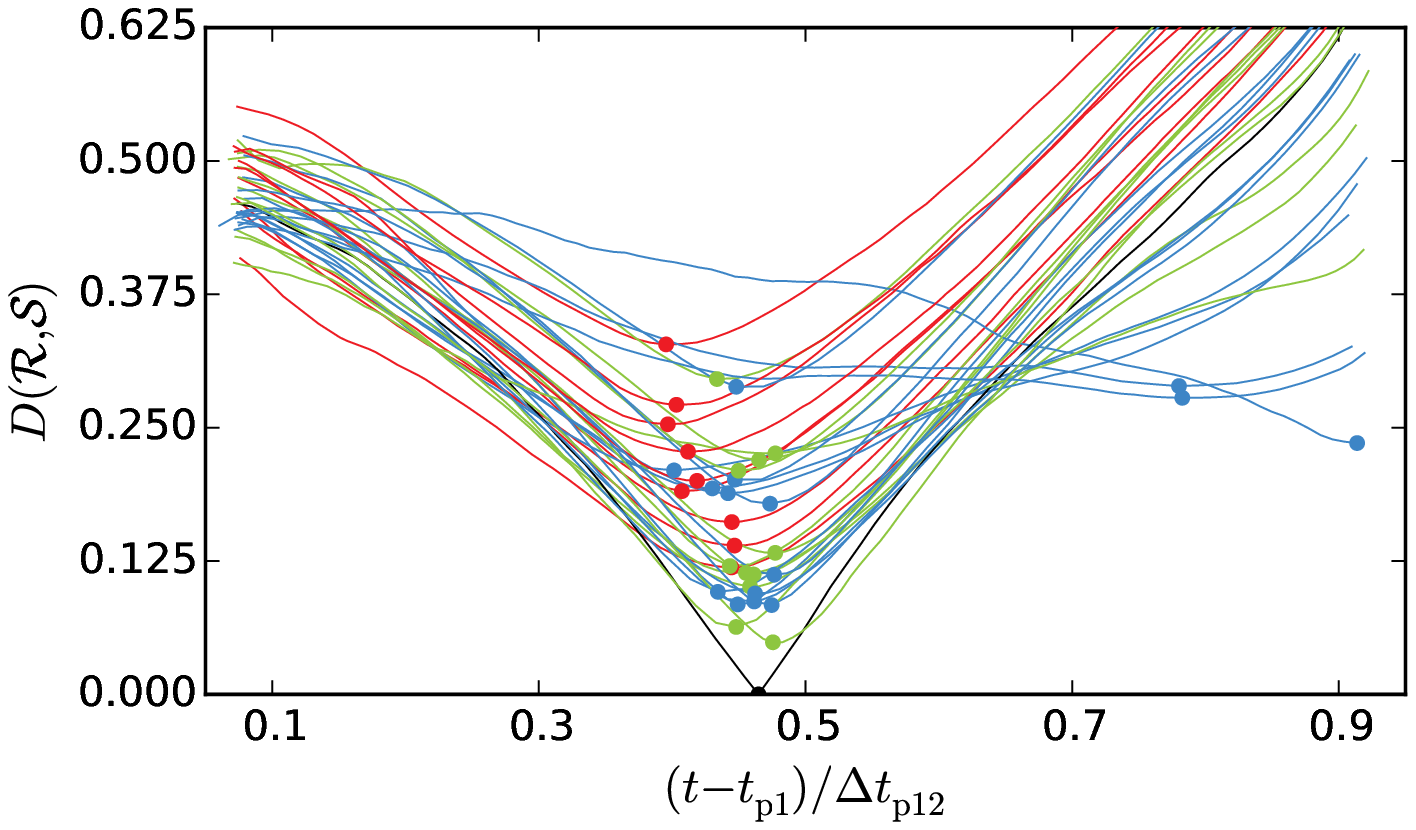,width=\columnwidth,
bbllx=76,bblly=263,bburx=480,bbury=498}
\epsfig{file=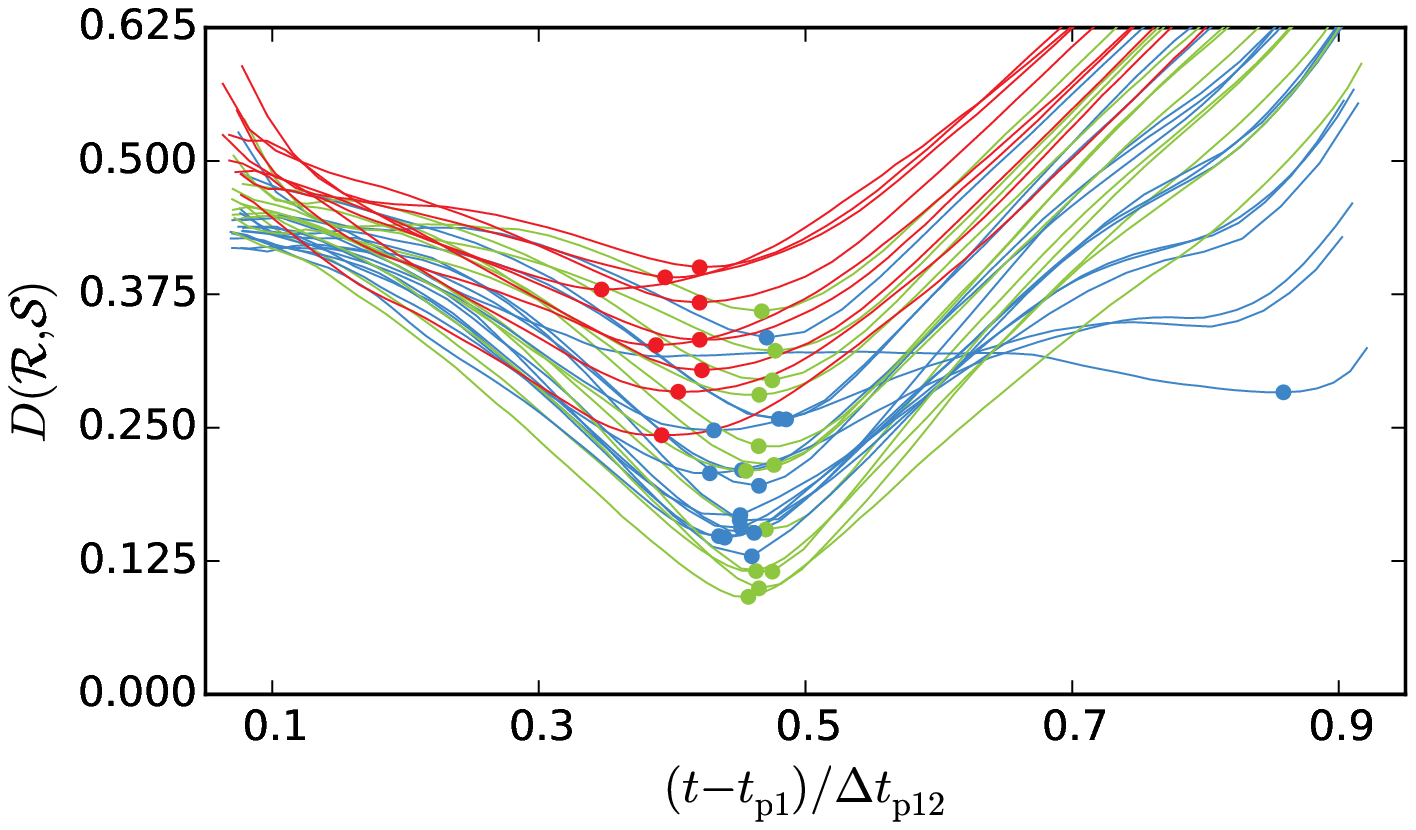,width=\columnwidth,
bbllx=76,bblly=263,bburx=480,bbury=498}
\epsfig{file=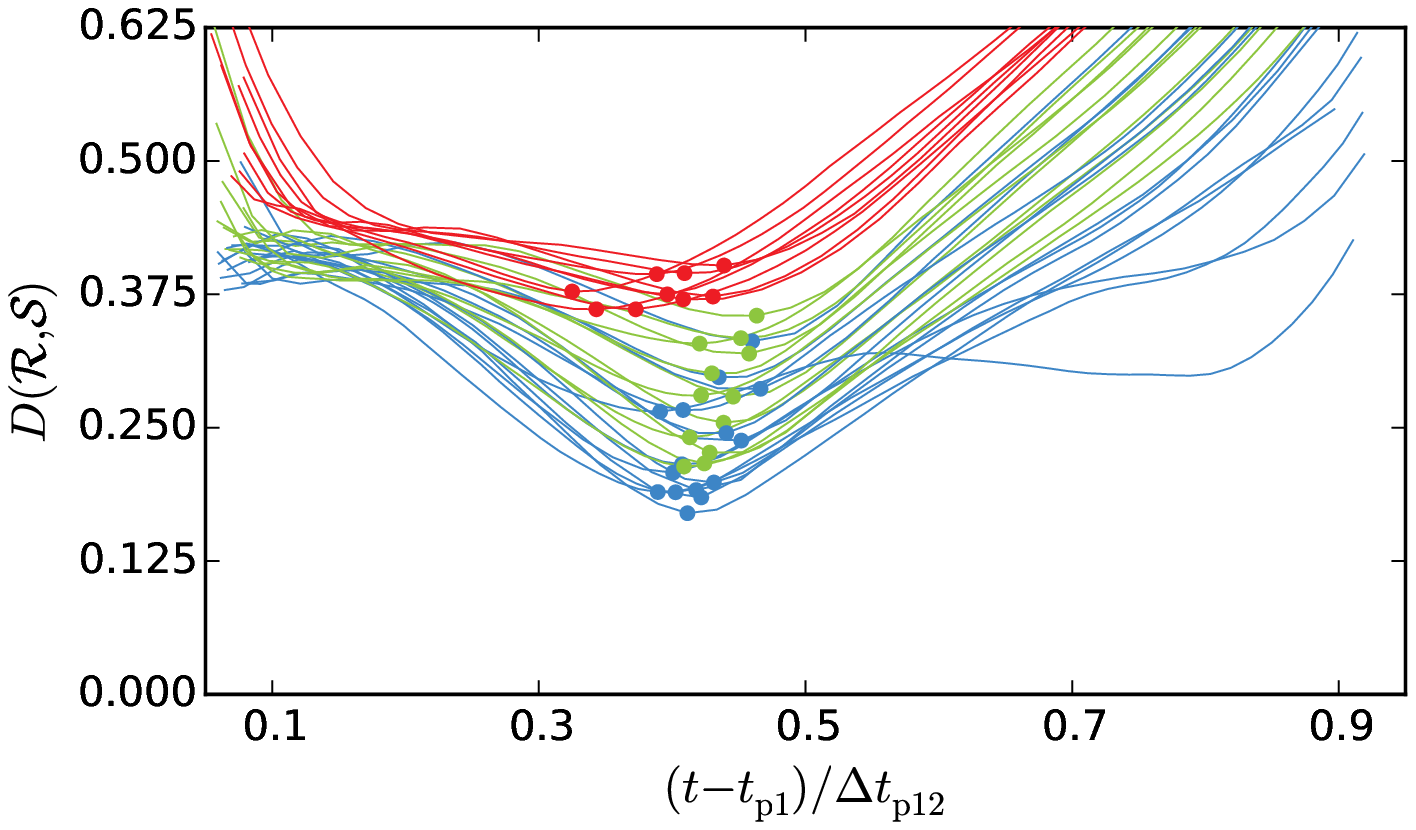,width=\columnwidth,
bbllx=76,bblly=263,bburx=480,bbury=498}
\caption{Time evolution of ${D}$ values \new{computed with respect to
the reference image $\mathcal{R}$ shown in the} lower left of
Fig.~\ref{fig:imgmatch}.  \new{The three panels show results for
comparison images $\mathcal{S}$ with $r_\peri{} / a_\halo = 2.0$
(top), $1.0$ (middle), and $0.5$ (bottom).}  Black curve:
\new{comparisons with reference encounter at other} times; \new{an
exact match}, ${D} = 0$, \new{occurs} at $t = t_\apo{1}$. Red, green,
blue curves show matches to other \new{encounters} with $f_\lumin =
0.2$, $0.1$, $0.05$, respectively.  \new{Minima on these curves,
corresponding to best matches with the reference image, are marked.}
\label{fig:radevol}}
\end{center}
\end{figure}

To compare tidal configurations, simulations are turned into images,
and differences are evaluated pixel by pixel.  Before this can be
done, some nuisance parameters must be dealt with.  Pixel comparison
will fail to recognize two geometrically similar shapes which don't
have the same scale, orientation, and position in the image plane
(Fig.~\ref{fig:imgmatch}, top).  Rather than blindly search for a
transformation which minimizes pixel differences, these parameters can
be eliminated by transforming the simulation data to register the
centres of the two galaxies at $\overline{\vect{r}}_1 = (1,0,0)$ and
$\overline{\vect{r}}_2 = (-1,0,0)$.  (Obviously, this is only possible
if the centres are well-separated; a different method is needed to
compare images of merger remnants.)  Next, the disc particles are
projected onto the image plane and adaptively smoothed to produce a
continuous grey-scale image which suppresses small-scale details but
captures the overall tidal structure.  To bring out tidal features,
which typically have low surface density, pixel values are
logarithmically transformed.  Finally, two such images, with pixel
values $\mathcal{A}_{k,l}$ and $\mathcal{B}_{k,l}$ where $k$ and $l$
are pixel indices, are compared by evaluating the normalized absolute
difference,
\begin{equation}
  {D}(\mathcal{A}, \mathcal{B}) =
    \frac{\sum_{k,l} |\mathcal{A}_{k,l} - \mathcal{B}_{k,l}|}
         {\sum_{k,l} \mathcal{A}_{k,l} + \mathcal{B}_{k,l}} \, .
  \label{eq:rad-def}
\end{equation}

Fig.~\ref{fig:imgmatch} presents two encounters which yield very
similar tidal configurations.  Here, the \new{encounter on the left,
which has} parameters $(f_\lumin, c_\halo, \alpha_\disc a_\halo,
r_\peri{} / a_\halo) = (0.1, 8, 3, 2)$, \new{was chosen as the
reference; it's} shown at first apocentre ($t_\mathrm{\new{ref}} =
t_\apo{1}$).  The \new{comparison encounter, in the middle, has
parameters} $(0.1, 16, 2.4, 2)$ \new{and} is shown \new{just} slightly
later; the \new{procedure used to select the encounter and time} will
be described shortly.  Both encounters are viewed perpendicular to the
orbital plane.  At the chosen times, the two galaxy pairs have
slightly different position angles and distinctly different
separations, and subtracting one image from another yields large
residuals (top right).  But these residuals are mostly due to
differences in orientation and scale; once the two galaxies have been
registered, the images are almost identical (bottom right).  Hence,
these two encounters are \new{highly} degenerate.

The match in Fig.~\ref{fig:imgmatch} was found by a simple linear
search.  First, \new{for every encounter $\mathcal{I}$}, a sequence of
registered images $\mathcal{S}_{\new{\mathcal{I}}}(t)$ spanning times
$t$ between $t_\peri{1}$ and $t_\peri{2}$ was generated and stored; on
average there are $\sim 70$ images per sequence.  \new{Let
$\mathcal{R}(t)$ be the sequence of images generated from the
reference encounter; the reference image, shown on the lower left in
Fig.~\ref{fig:imgmatch}, is $\mathcal{R}(t_\apo{1})$.  Next,
$\mathcal{R}(t_\apo{1})$ was compared to every other image, yielding
$D(\mathcal{R}(t_\apo{1}), \mathcal{S}_\mathcal{I}(t))$ values for
every encounter $\mathcal{I}$ and time $t$.  Finally, these values
were used to determine the time $t_\mathcal{I}$ when each
$\mathcal{S}_\mathcal{I}(t)$ best matches $\mathcal{R}(t_\apo{1})$,
and the corresponding $D(\mathcal{R}(t_\apo{1}),
\mathcal{S}_\mathcal{I}(t_\mathcal{I}))$ values were sorted to
identify the encounters most nearly degenerate with the reference
encounter.} The ${D}(\mathcal{R}(t_\apo{1}),
\mathcal{S}_{\new{\mathcal{I}}}(t))$ values are plotted as functions
of \new{$t$} in Fig.~\ref{fig:radevol}.  The black curve was obtained
by comparing $\mathcal{R}(t_\apo{1})$ with other images from the same
sequence, $\mathcal{R}(t)$; ${D}$ drops monotonically to zero when the
reference \new{image} is compared to itself, and subsequently
increases.~Curves in other colors show \new{the} results of comparing
\new{with} other encounters.

\new{While no other encounter prefectly matches the reference image},
in \new{most} cases ${D}$ reaches a definite minimum
${D}_\mathrm{min}$ at \new{nearly} the same stage between $t_\peri{1}$
and $t_\peri{2}$.  The timing of these minima is determined by the
time selected for the reference image.  The depth of each minimum
shows how closely the corresponding encounter \new{matches}
$\mathcal{R}(t_\apo{1})$.  \new{The deepest minima are found in the
top panel, which compares the reference image to images of other
encounters with the same pericentric separation, $r_\peri{} / a_\halo
= 2$.  Within the top panel,} most of the curves with deep minima are
produced by encounters with $f_\lumin = 0.1$ (green) or $f_\lumin =
0.05$ (blue), while those with $f_\lumin = 0.2$ (red) yield shallower
minima.  \new{This indicates that image matching can discriminate
between encounters with different luminous fractions or pericentric
separations.}  \new{In the top panel,} two curves have minima
significantly below the rest; the deeper one was used in
Fig.~\ref{fig:imgmatch}.  The reference encounter and its two closest
matches all have $f_\lumin = 0.1$ and $r_\peri{} / a_\halo = 2$.
\new{Moreover}, they have $(c_\halo, \alpha_\disc a_\halo)$ values of
$(16, 2.4)$, $(8, 3.0)$, and $(4, 3.75)$; \new{suggesting} that halo
concentration and disc scale can `trade off' against each other to
produce degenerate encounters.

Encounters which appear similar from one viewing direction don't
automatically look alike from another.  To investigate this effect,
registered images were constructed using four different viewing
directions $(\vect{u}_0, \vect{u}_1, \vect{u}_2, \vect{u}_3)$ derived
from the symmetry axes of a tetrahedron, with $\vect{u}_0$
perpendicular to the orbital plane.  The upshot, at least for these
encounters, is that ${D}_\mathrm{min}$ is fairly independent of
viewing direction, although $\vect{u}_0$ is typically the most
sensitive to differences in tidal configuration.  This is not
unexpected, since the same geometry is used for all the encounters.
It's convenient to use the average over all four directions,
$\overline{D}_\mathrm{min}$, as an overall measure of configuration
difference.  For example, the encounters in Fig.~\ref{fig:imgmatch}
match well along all four directions, with $(\vect{u}_0, \vect{u}_1,
\vect{u}_2, \vect{u}_3)$ yielding ${D}_\mathrm{min} = (0.049, 0.043,
0.046, 0.036)$, respectively; the average is
$\overline{D}_\mathrm{min} \simeq 0.044$.

\subsection{Patterns of degeneracy}

\begin{figure*}
\begin{center}
\epsfig{file=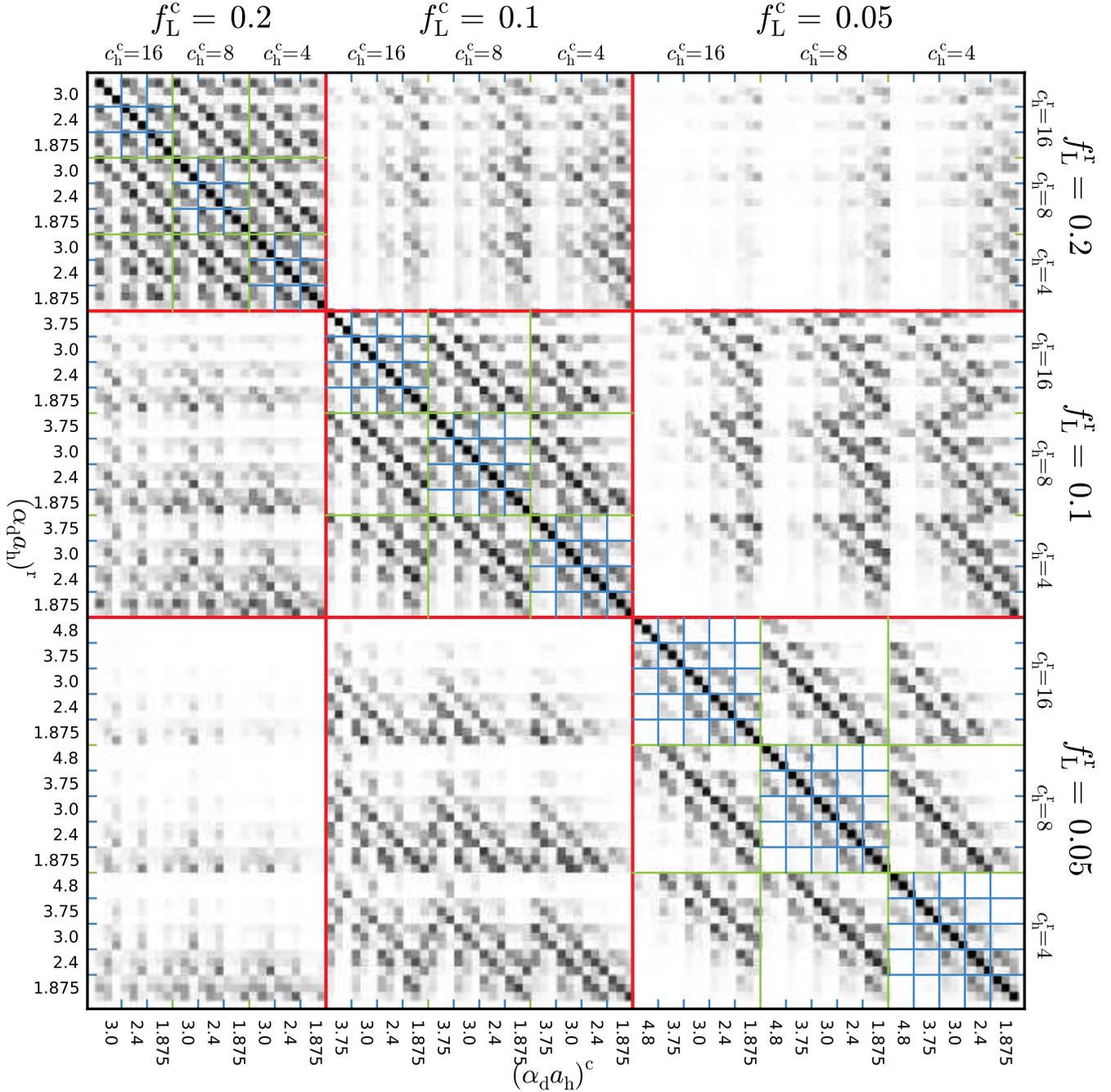,width=\textwidth,
bbllx=103,bblly=193,bburx=505,bbury=595}
\caption{$108 \times 108$ grid comparing tidal configurations.
\new{Each row of this grid shows the result of matching all $108$
comparison encounters to a given reference encounter. Encounters are
arranged in the \textit{same order} along both axes.} The red, green,
and blue lines separate encounters with different $f_\lumin$,
$c_\halo$, and $\alpha_\disc a_\halo$ values, respectively.  Values of
$f_\lumin$ and $c_\halo$ are given along the \new{top and right},
while $\alpha_\disc a_\halo$ is given on the \new{bottom and} left;
the `$\mathrm{r}$' and `$\mathrm{c}$' superscripts indicate reference
and comparison simulations, respectively.  In each $3 \times 3$ group
of cells, \new{$(r_\peri{} / a_\halo)^\mathrm{c}$ increases rightward,
and $(r_\peri{} / a_\halo)^\mathrm{r}$ increases downward}.  The grey
value for each cell is $1 - \exp(- \overline{D}_\mathrm{min}^2 /
0.01)$; this stretch emphasizes pairs which have similar
configurations.
\label{fig:cmpgrid}}
\end{center}
\end{figure*}

The same procedure has been applied to all $108$ encounters in the
standard ensemble.  In each case, a reference image generated at
apocentre, time $t_\mathrm{ref} = t_\apo{1}$, is compared to images
from all other simulations, and the minimum value
$\overline{D}_\mathrm{min}$ from each is used as a measure of
degeneracy.  Fig.~\ref{fig:cmpgrid}, \new{presents} the results,
\new{which summarize over three million image comparisons}.  Here
\new{encounters are arranged in the \textit{same order} along both
axes; along each axis, $r_\peri{} / a_\halo$ varies fastest,
$\alpha_\disc a_\halo$ next, $c_\halo$ next, and $f_\lumin$ slowest.}
\new{The} grey value of each cell indicates the degree of
\new{degeneracy}, ranging from black (identical) to white (different).
The diagonal line running from upper left to lower right shows that
each simulation perfectly matches itself, while dark off-diagonal
cells indicate encounters which are degenerate despite having
different parameters.  \new{Note that because a search over time is
done to match each comparison encounter to a given reference
encounter, this grid is not \textit{perfectly} symmetric about the
diagonal, although the asymmetry is rather subtle.}

\begin{figure}
\begin{center}
\epsfig{file=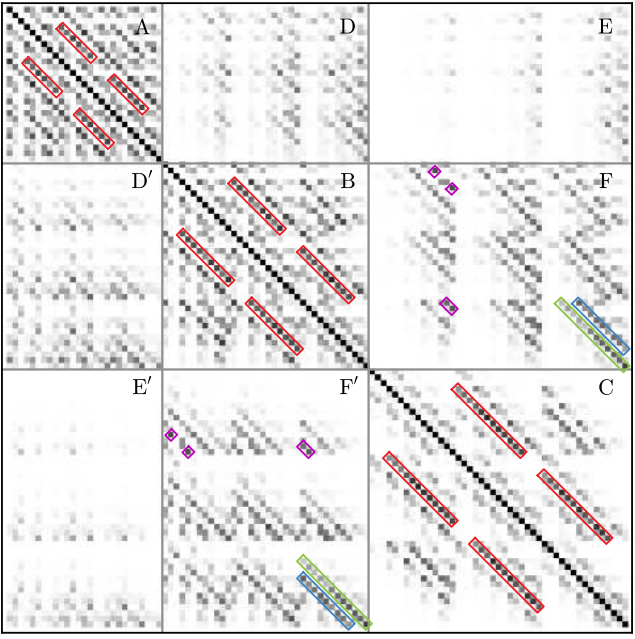,width=\columnwidth,
bbllx=132,bblly=222,bburx=480,bbury=570}
\caption{Key to Fig.~\ref{fig:cmpgrid}, identifying regions and
encounter sequences described in the text.
\label{fig:cmpgrid_key}}
\end{center}
\end{figure}

\begin{figure}
\begin{center}
\epsfig{file=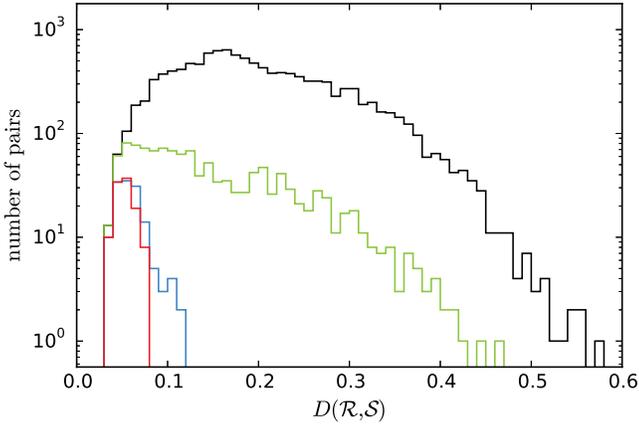,width=\columnwidth,
bbllx=87,bblly=246,bburx=489,bbury=512}
\caption{Distributions of $D$ values for subsets of the grid in
Fig.~\ref{fig:cmpgrid}.  \new{Cells on the main diagonal, which have
$D = 0$, are excluded from these subsets.}  The black histogram shows
values for all off-diagonal cells, while the green one shows cells
which have the same $f_\lumin$ and $r_\peri{} / a_\halo$.  The red and
blue histograms are described in the text.  Note the vertical axis is
logarithmic.
\label{fig:cmphist}}
\end{center}
\end{figure}

The first point Fig.~\ref{fig:cmpgrid} illustrates is that only a few
\new{pairs of} encounters are as degenerate as the \new{two} in
Fig.~\ref{fig:imgmatch}.  This implies that it may \new{indeed} be
possible to learn \new{something} about progenitor structure by
analyzing tidal configurations.  For example, \new{most of the close
matches are found in the three large squares recording comparisons
between encounters which have identical $f_\lumin$ values (labeled A,
B, and C in Fig.~\ref{fig:cmpgrid_key}).  Indeed, as
Fig.~\ref{fig:cmphist} shows, pairs of encounters with identical
values of $f_\lumin$ and $r_\peri{} / a_\halo$ account for most of the
smaller $D_\mathrm{min}$ values in Fig.~\ref{fig:cmpgrid}.  In
contrast, the almost complete absence of dark cells within the
rectangles labeled E and E' in Fig.~\ref{fig:cmpgrid_key} shows} that
encounters with $f_\lumin = 0.2$ and $0.05$ \new{produce distinctly}
different configurations, and \new{thus} are unlikely to be confused
with each other.

A second point is that degenerate pairs don't occur at random.
Fig.~\ref{fig:cmpgrid} exhibits a good deal of structure, with most of
the degenerate pairs \new{of encounters in regions A, B, and C
(Fig.~\ref{fig:cmpgrid_key})} arranged in sequences paralleling the
main diagonal.  The contrast between these sequences and their
surroundings is inversely correlated with luminous fraction, being
strongest when \new{the} encounters are halo-dominated, as in
\new{region C, where} $f_\lumin = 0.05$.

Perhaps the most obvious \new{of these} patterns are the diagonal
sequences associated with the trade off between halo concentration and
disc scale \new{noted in \S~\ref{sec:cmp_proc}}.  These degenerate
pairs of encounters always have the same $f_\lumin$ and $r_\peri{} /
a_\halo$, but differ in $c_\halo$ and $\alpha_\disc a_\halo$.  \new{In
Fig.~\ref{fig:cmpgrid_key}, they form the diagonal sequences outlined
in red.}  The \new{origin of these patterns} lies in the relationship
between halo concentration and apocentric distance (see
\S~\ref{sec:orbevol}): \new{in effect}, doubling $c_\halo$ increases
$r_\apo{1}$ by an average factor of $\sim 1.26 \pm 0.05$.  If the disc
scale is increased by the same factor, the resulting configuration
will closely match the original.  Fig.~\ref{fig:cmphist}, which plots
distributions of ${D}_\mathrm{min}$ for various sets of pairs, tests
this explanation.  Here the red histogram, \new{which corresponds to
the cells outlined in red in Fig.~\ref{fig:cmpgrid_key}, derives} from
pairs \new{of encounters} in which $c_\halo$ is doubled and
$\alpha_\disc a_\halo$ is scaled by a factor of $\sim 0.794 \simeq
1.26^{-1}$, while $f_\lumin$ and $r_\peri{} / a_\halo$ are held fixed.
This picks out a large fraction of the most degenerate \new{encounter
pairs}.  In comparison, the blue histogram shows the
${D}_\mathrm{min}$ distribution for pairs \new{of encounters} which
have similar $r_\apo{1} \alpha_\disc$ values\footnote{Specifically,
differing by no more than $0.035$ in $\log_{10}(r_\apo{1}
\alpha_\disc)$.}, with $f_\lumin$ and $r_\peri{} / a_\halo$ again
fixed.  This includes most of the pairs in the red histogram, as well
as a \new{few} additional pairs which are slightly less degenerate.

\new{Turning to encounters with different values of $f_\lumin$, the
strongest degeneracies arise when one encounter has $f_\lumin = 0.1$
and the other has $f_\lumin = 0.05$ (rectangles F and F' in
Fig.~\ref{fig:cmpgrid_key}).  Distinguishing between these cases is
observationally interesting, since many disc galaxies appear to have
$f_\lumin$ values in this range \citep{Z+2014}.  Some} of the diagonal
sequences \new{in rectangles F and F'} involve pairs with relatively
extended discs (ie, small $\alpha_\disc a_\halo$) and compact haloes
(small $c_\halo$).  As Fig.~\ref{fig:vrot} shows, the discs in these
models contribute \new{relatively} little to the total circular
velocity; in effect, these models are so halo-dominated that the discs
could almost be massless.  It's scarcely surprising that two such
encounters which have the same $c_\halo$, $\alpha_\disc a_\halo$, and
$r_\peri{} / a_\halo$ would yield similar configurations.  \new{For
example, the diagonal sequence outlined in green in
Fig.~\ref{fig:cmpgrid_key}} links encounters with \new{identical
$c_\halo$, $\alpha_\disc a_\halo$, and $r_\peri{} a_\halo$ values.
Notice that the lower right end of this sequence, which represents the
encounters with the most extended discs, shows the highest level of
degeneracy\footnote{\new{This includes the \textit{most} degenerate
pair of encounters in rectangles F and F', with parameters $(c_\halo,
\alpha_\disc a_\halo, r_\peri{} / a_\halo) = (4, 1.875, 2)$.  The
tidal features of the discs in these two encounters can be compared in
supplement Fig.~4 under the labels `4A6A' and `7A6A'.}}, while
encounters at the other end of the sequence are much more distinct.}

\new{Other sequences and individual matches in rectangles F and F' are
not so easily explained.  Some, including the sequence outlined in
blue in Fig.~\ref{fig:cmpgrid_key}, involve pairs of encounters with
similar values of $r_\apo{1} \alpha_\disc$, but this condition is
neither sufficient nor necessary.  A few moderately degenerate pairs
of encounters, like the ones outlined in magenta, have different
values of $r_\peri{} / a_\halo$.  Such pairings are not seen in other
regions of Fig.~\ref{fig:cmpgrid_key}; their presence here indicates
that it may not always be possible to constrain both $f_\lumin$ and
$r_\peri{} / a_\halo$ using tidal configuration.}

The pattern\new{s} seen in Fig.~\ref{fig:cmpgrid} \new{are} also found
when reference images are generated at other times (see supplement
Fig.~9).  In general, tidal configurations diverge with time, so
encounters are harder to distinguish before $t_\apo{1}$, and easier to
tell apart between $t_\apo{1}$ and $t_\peri{2}$.  Similar results are
also obtained when only one of the two discs is imaged (supplement
Fig.~10); in other words, the $i = 0^\circ$ and $i = 71^\circ$ discs
independently reproduce the patterns seen here.

\subsection{Morphological differences}

\begin{figure*}
\begin{center}
\epsfig{file=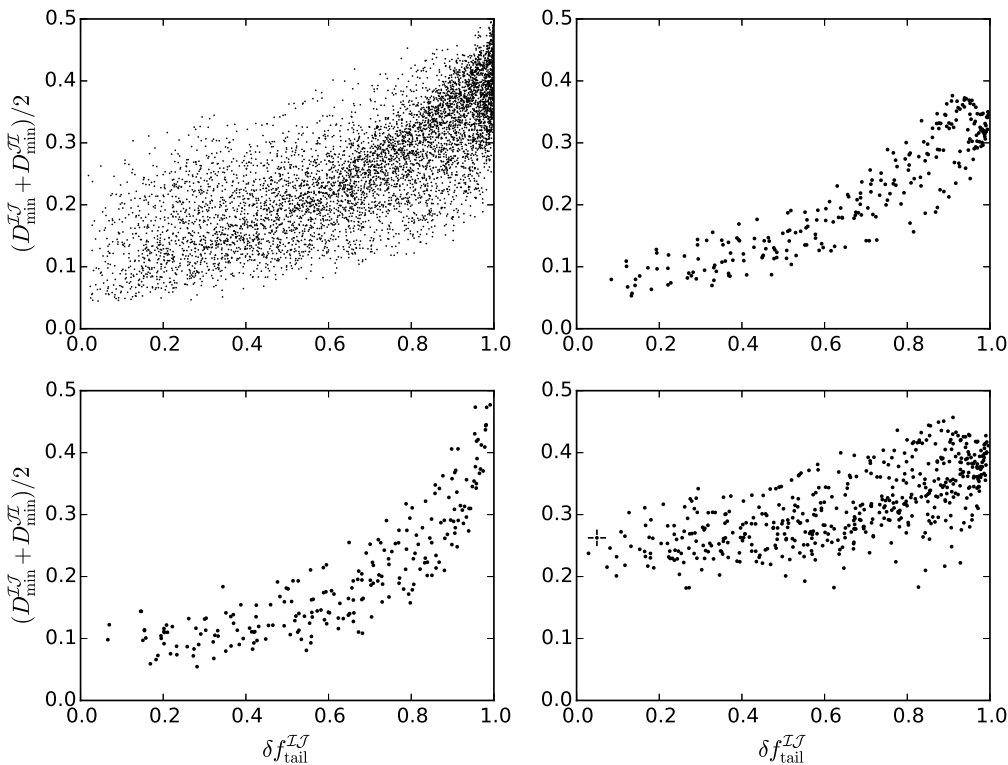,height=0.75\textwidth,angle=270,
bbllx=96,bblly=80,bburx=548,bbury=678}
\caption{Relationship between differences in tail fraction and
configuration.  The latter is symmetrized so that each pair of
encounters is represented by a single point.  Top left: result for
\textit{all} pairs of encounters.  Top right: pairs which undergo
violent orbit decay.  Bottom left: pairs which undergo gentle orbit
decay.  Bottom right: mixed pairs.  The pair marked with cross-hairs
is presented in Fig.~\ref{fig:morphdiff}.
\label{fig:radftail}}
\end{center}
\end{figure*}

Having some idea of the factors which make a pair of encounters
similar, it's logical to ask what makes them different.  One factor
which obviously plays a role is the amount of tidal material; other
things being equal, a larger fraction of tidal material increases the
surface brightness of extended features.  But does the tidal
\textit{morphology} also matter, or are the differences in
configuration measured by (\ref{eq:rad-def}) basically driven by
differences in tidal fraction?  If morphology matters, then it should
be possible to find \new{pairs of encounters}, with similar tidal
fractions, which \new{nonetheless} have visibly different
configurations.

Comparison of configurations is more effective after tidal features
have had time to develop, so this section will use a reference time
half-way between first apocentre and second pericentre:
$t_\mathrm{ref} = (t_\apo{1} + t_\peri{2}) / 2$.  At such late times,
tidal fractions have been substantially affected by reaccretion
(\S~\ref{sec:lifetail}).  Since a good measure of bridge reaccretion
is not yet available, tail fraction (Fig.~\ref{fig:tailevol}) will be
used as a proxy for overall tidal fraction.  To further improve the
discrimination of different configurations, only $\vect{u}_0$ images,
face-on to the orbital plane, will be used to compute
${D}_\mathrm{min}$.

Let $f_\tail{1}^\mathcal{I}$ and $f_\tail{2}^\mathcal{I}$ be tail
fractions for the two discs in encounter $\mathcal{I}$ at time
$t_\mathrm{ref}$.  A relative measure of the difference in tail
fractions for encounters $\mathcal{I}$ and $\mathcal{J}$ is
\begin{equation}
  \delta f_\tail{}^\mathcal{IJ} = 1 - 
    \min \!\! \left [
      \frac{\min(f_\tail{1}^\mathcal{I},
                 f_\tail{1}^\mathcal{J})}
           {\max(f_\tail{1}^\mathcal{I},
                 f_\tail{1}^\mathcal{J})} \, ,
      \frac{\min(f_\tail{2}^\mathcal{I},
                 f_\tail{2}^\mathcal{J})}
           {\max(f_\tail{2}^\mathcal{I},
                 f_\tail{2}^\mathcal{J})}
         \right ] \! .
  \label{eq:delta-ftail-def}
\end{equation}
This quantity vanishes if both tails in $\mathcal{I}$ are as massive
as their counterparts in $\mathcal{J}$, and increases if either
of the corresponding tail fractions are different.  Likewise, let
${D}_\mathrm{min}^\mathcal{IJ}$ be the minimum value of ${D}$ obtained
when matching the reference image of $\mathcal{I}$ against the
sequence of images of $\mathcal{J}$.  The upper left panel in
Fig.~\ref{fig:radftail} plots difference in configuration against
difference in tail fraction for all pairs of encounters in the
standard ensemble.  If configuration differences were largely driven
by tail fraction, then these parameters should be highly correlated,
and all pairs with $\delta f_\tail{} \simeq 0$ should have
${D}_\mathrm{min} \simeq 0$.  That's not what Fig.~\ref{fig:radftail}
shows; while ${D}_\mathrm{min}$ is correlated with $\delta f_\tail{}$,
it still spans a considerable range even for small $\delta f_\tail{}$.
Assuming tail fraction is a good proxy for overall response, it
appears that morphology does matter.

\begin{figure*}
\begin{center}
\begin{tabular}{cc}
\epsfig{file=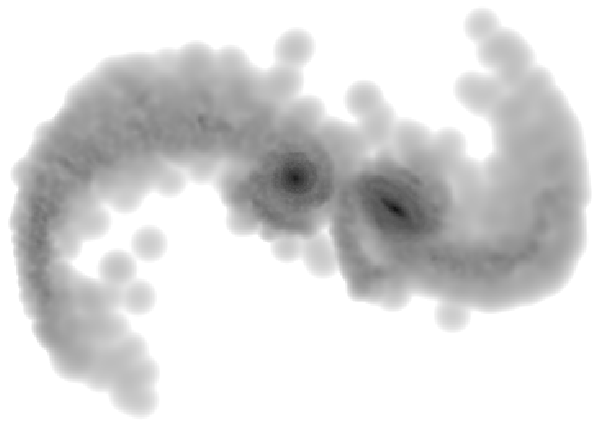,width=2.75in}&%
\epsfig{file=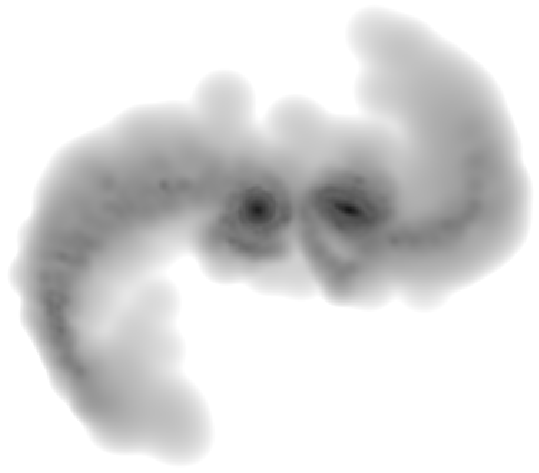,width=2.75in}\\
\epsfig{file=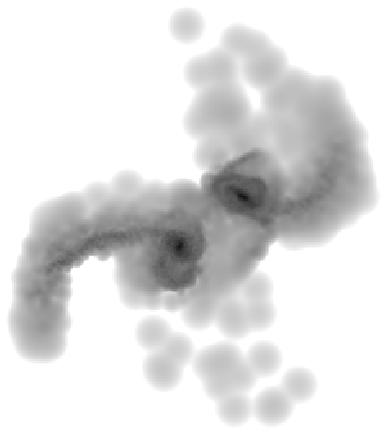,width=2.75in}&%
\epsfig{file=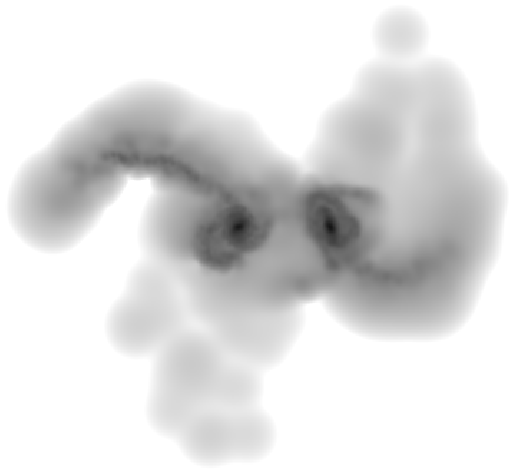,width=2.75in}\\
\end{tabular}
\caption{Two encounters with similar amounts of tail material at
$t = t_\mathrm{ref}$ but different morphologies.  Both are viewed
face-on to the orbit plane.  Top: encounter parameters $(f_\lumin,
c_\halo, \alpha_\disc a_\halo, r_\peri{} / a_\halo) = (0.2, 8, 2.4,
2)$.  Bottom: parameters $(0.1, 16, 3.75, 0.5)$.  Left: plotted to the
same scale.  Right: registered to $(\pm 1, 0, 0)$.
\label{fig:morphdiff}}
\end{center}
\end{figure*}

However, it is possible to identify subsets of the standard ensemble
where differences in tail fraction are more strongly correlated with
differences in configuration.  The upper right panel of
Fig.~\ref{fig:radftail} restricts the sample to pairs where both
encounters have orbits decaying `beyond the zero'
(\S~\ref{sec:orbevol}).  Specifically, the $23$ encounters in this
subset have orbital angular momenta at second pericentre $j_\peri{2} <
-0.05 \, j_\mathrm{orb}$.  In this case, a definite relationship
between $\delta f_\tail{}$ and ${D}_\mathrm{min}$ is evident, and
${D}_\mathrm{min}$ becomes fairly small as $\delta f_\tail{} \to 0$.
For this subset, it's plausible that differences in tail fraction
account for most of the measured differences in tidal configuration;
spot checks along the sequence reveal many pairs with similar shapes
(see supplement Fig.~11).  In other words, encounters which undergo
violent orbit decay appear to have relatively homogeneous tidal
morphologies over a wide range of tail fractions.

A similar result holds for encounters whose orbits decay gently.  The
lower left panel of Fig.~\ref{fig:radftail} shows pairs with orbital
angular momenta at second pericentre $j_\peri{2} > 0.15 \,
j_\mathrm{orb}$; this subset contains $22$ encounters.  Again, a
fairly well-defined relationship between $\delta f_\tail{}$ and
${D}_\mathrm{min}$ emerges, although in this case the relationship
becomes steeper as $\delta f_\tail{} \to 1$.  This may indicate that
these gentle orbit decays, as a group, are not quite so homogeneous,
but once again, spot checks show that many pairs have similar shapes
even though their tail fractions may differ (supplement Fig.~12).

Finally, the lower right panel of Fig.~\ref{fig:radftail} pairs
violent and gentle orbit decays; that is, one member of each pair has
$j_\peri{2} < -0.05 \, j_\mathrm{orb}$, while the other has
$j_\peri{2} > 0.15 \, j_\mathrm{orb}$.  Even when both members have
comparable amounts of tidal material, their morphologies are quite
different, yielding ${D}_\mathrm{min} \gtrsim 0.2$ for almost every
pair.  Moreover, unlike the two previous cases, there's only a weak
correlation between $\delta f_\tail{}$ and ${D}_\mathrm{min}$; for
this pair sample, differences in tail fraction don't have that much to
do with differences in configuration.

As an example, Fig.~\ref{fig:morphdiff} contrasts the effects of
gentle (top) and violent (bottom) orbit decay on tidal morphology.
These two encounters, represented by the marked point in the lower
right panel of Fig.~\ref{fig:radftail}, have very different
configurations even though their tail fractions are quite similar at
time $t_\mathrm{ref}$ ($f_\tail{1} = 0.0691 \pm 0.0017$ and
$f_\tail{2} = 0.0366 \pm 0.0008$).  The morphological differences seen
here arise in various ways.  For example, in the bottom images the
discs display well-developed loops of reaccreted tail material, while
such features are much less evident in the top images.  This follows
from the details of these two encounters.  On the bottom, a close
($r_\peri{} / a_\halo = 0.5$) encounter between galaxies with
relatively deep potential wells (\new{$\eta_\mathrm{esc} \simeq 0.627$,
or} $\mathcal{E} \simeq 5.08$) launched substantial tails which fell
back quickly to create the loops seen here.  On the top, a wider
($r_\peri{} / a_\halo = 2$) encounter between galaxies with shallower
wells (\new{$\eta_\mathrm{esc} \simeq 0.692$, or} $\mathcal{E} \simeq
4.18$) generated somewhat less massive but longer-lived tails which
don't form extensive loops of tidal material.

However, the most obvious difference in Fig.~\ref{fig:morphdiff} is
the shapes of the tails themselves.  The two galaxies on the top,
which barely grazed each other ($R_{1/2} / r_\peri{} \simeq 1.47$),
continue to orbit in a clockwise direction, while those on the bottom
interpenetrated deeply ($R_{1/2} / r_\peri{} \simeq 10.0$), reversed
direction, and are now approaching on a counter-clockwise trajectory.
These different orbital paths markedly influence the shapes of the
tidal tails, which distort so as to maintain continuity with the discs
which spawned them.  The tails in the top encounter describe great
sweeping arcs moving in the same clockwise direction as the galaxies
they came from.  In contrast, the the tails in the bottom encounter,
especially the one from $i = 0$ disc at left, have lost much of their
angular momentum to the same gravitational field which reversed the
orbital motions of their parent galaxies.  While their tips continue
to travel in a clockwise direction, these tails are predominantly
falling almost radially toward the centre of the system, and the
material nearest the galaxies is backtracking on plunging,
counter-clockwise orbits.

The connection between tail morphology and halo mass was noted by
\citet{MDH1998}, who report `[a]s we consider encounters involving
galaxies with increasing halo mass, the tails become straighter and
more anemic'.  Earlier, \citet{DMH1996} observed that tail morphology
is connected to orbital angular momentum, and the bottom rows of their
figs.~3 and~4 nicely illustrate the relationship between luminous
fraction and tail shape.  But these studies did not explicitly examine
the evolution of orbital trajectories, instead linking tail shape to
\textit{initial} orbital shape.  As the above quote indicates, they
also linked tail shape and tail mass; this was probably inevitable,
since the limited computing power at their disposal precluded the
extensive grid of models presented in this study.  Similar limitations
led \citet{DMH1999} to rely on test-particle simulations for the bulk
of their experiments; they used dynamical friction to implement orbit
decay, so their models could not reproduce the violent reversals of
orbital angular momentum described here.


\section{CONCLUSIONS}

This study examines the relationship between the internal structure of
interacting disc galaxies on the one hand and the their orbital
dynamics and tidal morphology on the other.  \new{This has been
accomplished by constructing} a grid of bulge/disc/halo galaxy models;
these models include a subset broadly consistent with $\Lambda$CDM
predictions for bright disc galaxies \citep{MMW1998}, as well as
others with higher luminous fractions and proto-galactic spin
parameters.  Simulated encounters between identical models display a
range of \new{outcomes}, which are analyzed to investigate orbital
evolution, level of tidal response, and comparative tidal morphology.

Orbit decay, which is largely mediated by dynamical interactions
between massive dark haloes, follows a consistent pattern for all of
these equal-mass encounters.  The ratio $R_{1/2} / r_\peri{}$, which
compares the galactic half-mass radius to the pericentric distance of
the initial orbit, predicts many aspects of orbital evolution.  These
include the time-scale for orbit decay (Fig.~\ref{fig:decaytime}) and
the actual pericentric separation (Fig.~\ref{fig:interpen}).  This
ratio also correlates with the shape of the post-encounter orbits;
deeply interpenetrating encounters ($R_{1/2} / r_\peri{} \gtrsim 5$)
of galaxies with dominant haloes ($f_\lumin \le 0.1$) reverse
direction after first passage and follow self-crossing trajectories
due to violent transfer of orbital angular momentum to dark haloes.

Strength of tidal response is measured by a simple, consistent method,
and a straightforward \new{algorithm is defined} to distinguish
bridges and tails (Fig.~\ref{fig:tdclexamp}).  \new{This method works}
well for the two disc orientations ($i_1 = 0^\circ$ and $i_2 =
71^\circ$, $\omega_2^\mathrm{eff} \simeq 0^\circ$) featured in this
study, and may be useful in other cases as well.  SW's result that
tidal response correlates with the escape parameter $\mathcal{E}$,
defined in (\ref{eq:sw-escape-param}), is strongly confirmed.  In
particular, once tail and bridge responses can be separately measured,
$f_\tail{}$ displays an \new{remarkably} linear relationship with
\new{$\eta_\mathrm{esc}$ (equivalently, with $\mathcal{E}^{-1/2}$)};
pericentric separation $r_\peri{} / a_\halo$ and inclination $i$
control the slope and intercept (Fig.~\ref{fig:tidalfrac}).  Tidal
features have finite lifetimes before they're reaccreted by their
parent galaxies, and galaxies with larger $\mathcal{E}$ values
reabsorb their tails faster (Figs.~\ref{fig:tailevol} and
\ref{fig:tailfrac}).

Overall distribution of tidal material depends on many factors,
including orbit decay, tidal response, and rate of reaccretion.  A
systematic comparison of tidal configurations shows that some
encounters have very similar morphologies and would be difficult to
distinguish observationally (Fig.~\ref{fig:imgmatch}).  For example,
halo concentration and disc compactness are partly degenerate; a small
change in the latter can mask a large change in the former.  On the
other hand, variations in total luminous fraction $f_\lumin$ have
definite effects on tidal configuration which aren't easily masked by
changes in other parameters.  In particular, the violent orbit decays
characteristic of close encounters between massive, extended haloes
yield distinctive tidal morphologies quite different from those
produced in encounters between low-mass haloes
(Fig.~\ref{fig:morphdiff}).

This result may seem at odds with \citet{MDH1998}, who found that `it
may be difficult to distinguish between close collisions of low-mass
models and wider collisions involving more massive galaxies'.
However, they focused on models of the merger \textit{remnant}
NGC~7252, whereas the emphasis here is on tidal morphology between
first and second passage.  It's likely that the earlier dynamical
stage considered here provides more leverage on encounter parameters,
including pericentric separation; in reconstructing tidal encounters
from morphological and kinematic data, the separation and orientation
of still-distinct galaxies provides useful constraints \citep{BH2009}.

In examining morphological indicators which yield information on halo
structure, this study does not explicitly include kinematic data.  Of
course, line-of-sight velocities are necessary to fix an overall mass
scale, and they play a key role in accurately constraining the
encounter and viewing geometry of tidally interacting galaxies.
However, it's by no means clear that kinematic data would break the
degeneracies noted above; the configurations and velocity fields of
tidal features are intimately related, so the latter may not provide
much additional information about halo properties.

The results presented here don't necessarily create tension between
observations of long-tailed interacting galaxies or twin-tailed merger
remnants on the one hand, and the predictions of $\Lambda$CDM on the
other.  As long as a good-sized subset of the models in
Fig.~\ref{fig:vrot} have analogs among real galaxies, some fraction of
encounters will inevitably produce objects with `classic' tidal
features; this statement stands even if the $f_\lumin = 0.2$ models
are excluded.  The main requirement is that at least some galactic
discs extend far enough to allow tidal tails to develop and -- more
importantly -- persist after the galaxies merge.

The sizes of galactic discs predicted in $\Lambda$CDM are a bit
uncertain.  \citet{MMW1998} included a parameter specifying the
fraction of angular momentum retained by baryons as they form a disc,
and found that this must be near unity to match the observations.
Numerical simulations initially produced discs which were much too
small \citep[e.g.,][]{KG1991}.  However, the simulation results are
sensitive to the method used to model the gas, with recent moving-mesh
codes producing discs both larger and more organized than those
produced by SPH codes \citep{KVSSH2012, TVSSH2012}.  \new{Baryon
physics may} have a significant influence on the amount of angular
momentum galactic discs \new{acquire and retain \citep{B+2011,
SBBMDWM2013, G+2015}}.

Long-tailed merger remnants such as NGC~7252 \citep{S1982} appear to
place general constraints on progenitor models.  For $f_\lumin \simeq
0.1$, \new{remnants with prominent tails} can probably be produced
using a substantial range of disc scales, but if $f_\lumin \simeq
0.05$ galaxies are used, their discs must be very extended
($\alpha_\disc a_\halo \le 2.0$), in general agreement with earlier
results by \citet{MDH1998}.  There are several loopholes which may
weaken these constraints.  First, encounters closer than those
considered here may (a) increase the amount of tail material initially
launched, and (b) reduce the merger time-scale, allowing more of this
tail material to linger after the participants merge.  Further
experiments to investigate this possibility are warranted.  Second,
galaxies have neutral-hydrogen discs extending beyond their stellar
counterparts, and initially gaseous tails might be lit up by
interaction-induced star formation, producing optically prominent
tidal features even after the stellar tails have been reaccreted.
This scenario requires high rates of star formation \citep{MDH1998}
which seem at odds with the typical colors of tidal tails
\citep{SWSM1990, SGSH2010}.  Nonetheless, it may be worth estimating
the stellar masses of tidal features via multi-band photometric
methods to better constrain the fraction of old stellar material they
actually contain.

The main point of this study is that interacting disc galaxies are
likely to display a variety of tidal features due to differences in
progenitor structure.  For example, it seems likely that some deeply
interpenetrating encounters will \new{drive} orbital angular momentum
`beyond the zero' after first passage.  In such encounters,
low-inclination discs develop rather linear tails which extend outward
along the nearly-radial trajectories the galaxies follow after their
first passage (Fig.~\ref{fig:morphdiff}, bottom), instead of the
grandly curving tails first described by TT.  Recognizing such objects
may not always be trivial since tails can also appear linear when
viewed edge-on.  Arp~238 \citep{A1966} could be one instance; the
fainter tail to the South-East shows a roughly linear form, but
spatially-resolved velocity data is needed to substantiate the
impression that the orbit plane of this system is roughly
perpendicular to our line of sight.

As another example, encounters \new{between} galaxies with
\new{$\eta_\mathrm{esc} \lesssim 0.58$ (i.e., $\mathcal{E} \gtrsim
6.0$)} produce short-lived tails which are largely reaccreted before
second pericentre (Fig.~\ref{fig:tailfrac}).  \new{If observed between
the first and second passages}, such a system \new{may look like} a
pair of peculiar spirals \new{without obvious signs} of interaction.
After merging, the remnant may \new{appear disturbed}, yet lack the
long tidal tails which \new{signal a merger} between disc galaxies.
This scenario might explain certain enigmatic objects.  Arp~220, for
instance, is almost certainly the remnant of a merger between two
gas-rich disc galaxies \citep{S+1998}, \new{but does not display} the
conspicuous tails of NGC~7252.  While the absence of tails might be
explained by \new{the} encounter and viewing geometry, it's \new{also}
possible that this system reaccreted its tails before merging,
producing a confused and partly phase-mixed object.

\new{These results have interesting implications for estimates of
merger rates both locally and as a function of redshift.  Locally,
plausible variations in disc scale and halo structure from galaxy to
galaxy imply that some encounters will display conspicuous and
long-lived tidal features, while others, even under the most favorable
circumstances, may only briefly be recognizable as merging systems
(Figs.~\ref{fig:tidalfrac} and \ref{fig:tailfrac}).  Thus, samples
selected on the basis of optical morphology systematically
over-represent systems with unusually high luminous fractions and/or
extended discs, and under-count mergers between galaxies with deep
potential wells.  Moreover, if galaxy discs grow from the inside out,
mergers at high redshift will be consistently harder to detect via
morphology, even \textit{after} due allowance has been made for
band-shifting, cosmological dimming, and resolution effects.  In other
words, estimates of the `observability' time $\langle
T_\mathrm{obs}(z) \rangle$ \citep{LJCCPSS2011} based on models of
low-redshift galaxies could systematically overestimate $\langle
T_\mathrm{obs}(z) \rangle$ at high redshifts.}

Detailed modeling of individual systems, observed between first and
second passage and matching both morphology and line-of-sight velocity
data, appears to have a good chance of constraining the luminous
fractions of the progenitor disc galaxies to somewhat better than a
factor of~$2$.  Systematic modeling efforts, using a variety of
progenitor galaxy models, can be undertaken to test this.  It may be
misleading to focus exclusively on `textbook' examples of tidal
interactions.  Samples selected using, e.g., infrared luminosity
\citep{A+2009}, may better reflect the full range of progenitor
structures.

A further reason to undertake such modeling is to test the
\textit{dynamical} nature of the dark matter.  Most probes of dark
matter on galactic scales, including rotation curves, satellite
kinematics, and weak lensing, basically measure the gravitational
potential in a static situation, and infer the density of dark matter
using Poisson's equation.  This inference may be incorrect.  For
example, in MOND the relationship between potential and density
diverges from Poisson's equation in the low-acceleration limit
\citep{BM1984, SM2002}.  Tidal features presumably evolve in the
low-acceleration regime, so MOND may be just as effective as dark
matter at limiting the length and mass of tidal tails.  On the other
hand, without an unseen sink for angular momentum, the orbits of
interacting galaxies are expected to decay rather gradually in MOND
\citep{TC2008}; in particular, the violent orbit decays seen here are
precluded.  If modeling of interacting systems provides clear evidence
that \textit{momentum} is being transferred from luminous material to
an unseen component, we would have a fundamental reason to think that
the dark stuff really is matter.

I thank colleagues at the Institute for Astronomy, the Yukawa
Institute for Theoretical Physics, the Tokyo Institute of Technology,
Kyoto University, \new{and the RIKEN Advanced Institute for
Computational Science} for comments and suggestions.  I am grateful to
Misao Sasaki and Atsushi Taruya of the Yukawa Institute for their
hospitality.  \new{Kelly Blumenthal}, Lars Hernquist, Chris Mihos,
\new{George Privon, and Volker Springel provided helpful feedback on
earlier drafts of this paper.  Finally, I thank John Hibbard for a
very comprehensive referee's report which caught several mistakes and
helped me clarify the presentation.}



\section*{APPENDIX A: TECHNICAL DETAILS}

Within the body of the paper all results are quoted as dimensionless
ratios, which may easily be scaled to any particular physical
situation.  However, when describing the simulations it's convenient
to use an explicit system of units.  This system is defined by setting
the gravitational constant $G = 1$, the halo length scale to $a_\halo
= 0.25$, and the luminous mass to $M_\lumin = 0.25$.  With these
values fixed, all other parameters can be derived.

\paragraph*{Halo.}  The halo mass is $M_\halo = (f_\lumin^{-1} - 1) \,
M_\lumin$ and the halo taper radius is $b_\halo = c_\halo \, a_\halo$.
Table~\ref{table:halo_params} lists these parameters, along with the
halo masses enclosed within $a_\halo$ and $b_\halo$.

\begin{table*}
\begin{tabular}{l|ccc|ccc|ccc}
          & \multicolumn{3}{c|}{$f_\lumin = 0.2$} 
          & \multicolumn{3}{c|}{$f_\lumin = 0.1$} 
          & \multicolumn{3}{c}{$f_\lumin = 0.05$} \\
$c_\halo$ & 16 & 8 & 4 & 16 & 8 & 4 & 16 & 8 & 4 \\
\hline
$M_\halo$ & 1.0 & 1.0 & 1.0 & 2.25 & 2.25 & 2.25 & 4.75 & 4.75 & 4.75 \\
$b_\halo$ & 4.0 & 2.0 & 1.0 & 4.0 & 2.0 & 1.0 & 4.0 & 2.0 & 1.0 \\
$M_\halo(a_\halo)$ & 0.0895 & 0.1172 & 0.1586 & 0.2014 & 0.2638 &
0.3568 & 0.4251 & 0.5569 & 0.7532 \\
$M_\halo(b_\halo)$ & 0.8767 & 0.7941 & 0.6645 & 1.9725 & 1.7868 &
1.4951 & 4.1642 & 3.7722 & 3.1563 \\
\end{tabular}
\caption{Halo parameter values.
\label{table:halo_params}}
\end{table*}

\paragraph*{Disc.}  The disc mass is $M_\disc = 0.75 \,
M_\lumin = 0.1875$.  The inverse disc scale radius is $\alpha_\disc =
(7.5, 9.6, 12.0, 15.0, 19.2, 24.0)$ for $\alpha_\disc a_\halo =
(1.875, 2.4, 3.0, 3.75, 4.8, 6.0)$, respectively (models with
$\alpha_\disc a_\halo = 6.0$ were not used in the survey since they
are bar unstable).  Vertical scale heights $z_\disc = 0.125 /
\alpha_\disc$ range from $1/60$ to $1/192$ for these models.

\paragraph*{Bulge.}  The bulge mass is $M_\bulge = 0.25 \,
M_\lumin = 0.0625$.  The bulge scale length is $a_\bulge = 0.16 \,
a_\halo = 0.04$.

\vskip 6pt

In the $N$-body simulations, gravitational forces were `softened'
\citep{A1963} by smoothing each body with a \citet{P1911} kernel of
radius $\epsilon = 0.01 a_\halo = 0.0025$ before calculating the
Newtonian potential \citep{B2012}.  This choice implies that the
gravitational field of the inner $r^{-1}$ profile of the NFW model
halo is well-resolved down to a scale of $\sim 0.03 a_\halo$.  The
field of the inner $r^{-2}$ profile of the \citeauthor{J1983} model
bulge is poorly resolved, since $\epsilon = a_\bulge / 16$, but the
internal dynamics of the bulges are not critical for these
experiments.  For both of these components, accurate gravitational
fields are calculated using the smoothing formalism \citep{B2012}.

The disc field is always well-resolved in the radial direction
($\epsilon \alpha_\disc \le 0.048$ for all models with $\alpha_\disc
\le 19.2$), but somewhat less so in the vertical direction.  When
initializing the bulge and halo, the effects of softening on the disc
field are approximated by an interpolation function
\citep[eq.~A1]{BH2009}, with parameters determined by numerical
fitting.

$N$-body realizations were generated as described in
\S~\ref{sec:init}.  Bodies representing the halo are $4$ times as
massive as those representing the luminous components; this provides
better sampling of the latter at a modest computational cost.  

All encounters start with the two galaxies at an initial separation
$r_\mathrm{init} > 5$ length units, or $20 a_\halo$.  This reflects a
compromise between opposing considerations.  On the one hand, closer
starts take less time to reach pericentre, reducing both CPU time and
numerical relaxation.  On the other hand, if the galaxies initially
overlap to any significant degree then a Keplerian orbit won't
accurately specify their initial positions and velocities.  At
$r_\mathrm{init} = 5$, overlap is effectively negligible for models
with $c_\halo \le 8$; the halo mass has converged to $\sim
0.04$~percent by $R = r_\mathrm{init}$.  For the $c_\halo = 16$
models, $\sim 4$~percent of the halo mass lies outside $R =
r_\mathrm{init}$.  However, this has little effect on the net
potential energy of the initial configuration; the implied offset in
the initial velocities is $\sim 1$~percent.

\begin{figure*}
\begin{center}
\epsfig{file=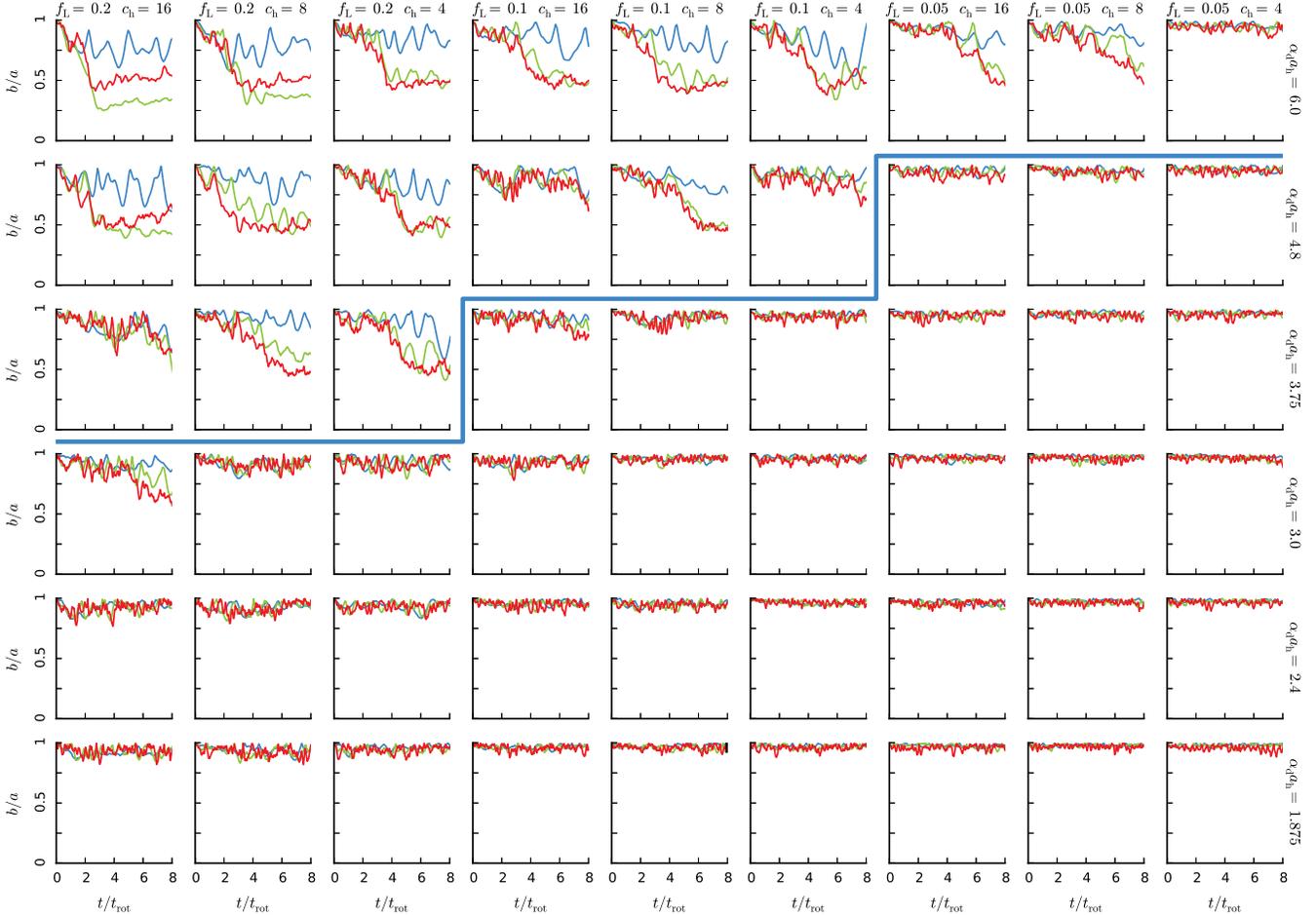,height=\textwidth,angle=270,
bbllx=107,bblly=105,bburx=518,bbury=690}
\caption{Stability tests for bulge/disc/halo galaxy models.  The
layout is similar to Fig.~\ref{fig:vrot} but \new{adds another} row of
models with $\alpha_\disc a_\halo = 6.0$ at the top; \new{the heavy
blue line is the same approximate stability boundary shown
previously.}  Red, green, and blue curves show $b/a$ ratios for the
inner $25$~percent, $50$~percent and $75$~percent of each disc,
respectively.
\label{fig:bar}}
\end{center}
\end{figure*}

Simulations were run using a hierarchical $N$-body code\footnote{For a
description of the algorithm, please see
\url{http://www.ifa.hawaii.edu/~barnes/treecode/treeguide.html}.}.
An opening angle of $\theta = 0.8$, together with quadrapole moments,
yields accelerations with median errors $\delta\vect{a}/|\vect{a}|
\lesssim 0.0006$.  Trajectories were integrated using a time-centred
leap-frog, with a time-step $\Delta t = 1/1024$ for all bodies.

\section*{APPENDIX B: STABILITY TESTS}

Galaxy models with massive, largely self-gravitating discs tend to be
dynamically unstable to bar formation.  In the context of the present
study, instability results when the parameter $\alpha_\disc a_\halo$
exceeds a critical value which depends on the chosen values of
$f_\lumin$ and $c_\halo$.  The rationale for excluding unstable models
from these experiments is complex.  Violently unstable initial models
are unrealistic.  There's no plausible way that nature could assemble
such configurations, since a bar would develop long before the entire
mass of the galaxy was in place; presumably the bar would then heat
the disc, producing a near-equilibrium configuration.  Starting with a
violently unstable system and allowing it to evolve is unlikely to
produce a similar configuration.  Mildly unstable initial conditions
are more plausible, provided the bars have enough time to develop and
approach equilibrium before the galaxies interact.  Tidal interactions
of barred galaxies are certainly interesting \citep[e.g.,][]{GCA1990},
but introduce yet another parameter -- the phase angle of the bar at
the time of pericentre -- into a problem which already has a daunting
number of parameters.  For practical reasons, such encounters are
excluded from the present study.

To determine the stability boundary in Fig.~\ref{fig:vrot}, individual
galaxy models were set up following the procedure in \S~\ref{sec:init}
and run in isolation.  Visually reviewing the resulting array of
simulations proved quite exhausting, so a simple procedure was
developed to quantify the (in)stability of the models.  At the start
of each simulation, disc bodies are sorted by orbital radius.  This
makes it easy to track the Lagrangian volumes corresponding to the
inner $25$~percent, $50$~percent and $75$~percent of each disc.  At
each time step, the moment-of-inertia tensor for each of these volumes
is computed, its eigenvalues $q_1$, $q_2$ are determined, and the
axial ratio $b/a = \sqrt{\min(q_1, q_2) / \max(q_1, q_2)}$ is
evaluated.  These axial ratios are plotted as functions of time in
Fig.~\ref{fig:bar}.  Each model was run to time $t = 8
t_\mathrm{rot}$, where $t_\mathrm{rot}$ is the rotation period at $R =
2 \alpha_\disc^{-1}$, to insure bars have time to develop.

Models \new{shown} in the bottom rows of Fig.~\ref{fig:bar} maintain
axial ratios $b/a \simeq 1$ for the duration of the simulations.
These discs exhibit transitory spiral features, but no lasting
bisymmetric patterns emerge.  On the other hand, disc-dominated
models, such as those in the upper-left region of this diagram, are
violently unstable; within a few $t_\mathrm{rot}$ they develop strong
bars within the inner $50$~percent of their discs.  For the purposes
of the present study, I adopted the stability boundary shown in
Figs.~\ref{fig:vrot} \new{and~\ref{fig:bar}}, fixing the maximum value
of $\alpha_\disc a_\halo$ for a given value of $f_\lumin$ independent
of $c_\halo$.  This choice is somewhat arbitrary; it included the
models with $(f_\lumin, c_\halo, \alpha_\disc a_\halo) = (0.2, 16,
3.0)$ and $(0.1, 16, 3.75)$, which both seem to be developing small
bars at later times, and excluded the model with $(f_\lumin, c_\halo,
\alpha_\disc a_\halo) = (0.05, 4, 6.0)$, which appears to be stable.
However, the regular structure of the adopted boundary makes the
organization and presentation of the experiments much more
straightforward.

\begin{figure}
\begin{center}
\epsfig{file=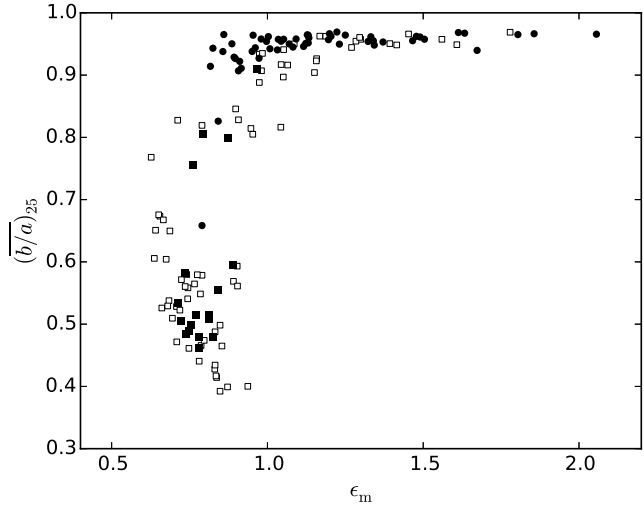,height=\columnwidth,angle=270,
bbllx=165,bblly=179,bburx=474,bbury=570}
\caption{Scatter plot of the axial ratio for the inner disc quartile,
$(\overline{b/a})_\mathrm{25}$, time-averaged for $6 \le
t/t_\mathrm{rot} \le 8$, versus the stability index
$\epsilon_\mathrm{m}$.  Open squares are models without bulges (i.e.,
the disc mass is increased by a factor of $1/3$).  Filled circles are
models below the stability boundary in Fig.~\ref{fig:vrot}, while
filled squares are above the stability boundary.  The $\alpha_\disc
a_\halo$ values represented range from $6.0$ to~$1.2$.
\label{fig:epsmbar}}
\end{center}
\end{figure}

While a general stability criterion is not easy to define, stable disc
models typically satisfy $\epsilon_\mathrm{m} \gtrsim
\epsilon_\mathrm{m,crit}$, where
\begin{equation}
  \epsilon_\mathrm{m} =
    \frac{v_\mathrm{max}}{\sqrt{G M_\disc \alpha_\disc}} \, ,
  \label{eq:epsm-def}
\end{equation}
$v_\mathrm{max}$ is the maximum rotation velocity of the model, and
the stability threshold is in the range $\epsilon_\mathrm{m,crit}
\simeq 1.1$ \citep{ELN1982} to $\epsilon_\mathrm{m,crit} \simeq 0.75$
\citep{SMM1998}.  Fig.~\ref{fig:epsmbar} plots the relationship
between $\epsilon_\mathrm{m}$ and the time-averaged value of the $b/a$
ratio for the inner quartile of each disc; the average is taken over
times $6 \le t/t_\mathrm{rot} \le 8$.  Open symbols represent a
parallel set of models without bulges; in these models, disc masses
are increased by factors of $1/3$, so as to obtain the same
$f_\lumin$, and all other parameters are left unchanged.  It appears
that $\epsilon_\mathrm{m}$ provides a approximate criterion for
stability, but doesn't always characterize marginally stable discs
correctly.  In particular, bulges exert a stabilizing influence which
is not completely reflected in the value of $\epsilon_\mathrm{m}$.

\end{document}